\documentclass[%
 reprint,
superscriptaddress,
 amsmath,amssymb,
 aip, 
 jcp, 
floatfix,
longbibliography
]{revtex4-2}

\usepackage{graphicx}
\usepackage{dcolumn}
\usepackage{bm}
\usepackage{hyperref}
\usepackage[normalem]{ulem}

\usepackage{xcolor} 
\usepackage{pdfpages}
\usepackage{pgffor}

\usepackage{siunitx}

\DeclareSIUnit{\angstrom}{\text{\AA}}
\DeclareSIUnit{\calorie}{\text{cal}}
\newcommand{\kB}{k_\textrm{B}}
\newcommand{\dd}{\mathrm{d}}

\newcommand{\ilm}{Universit{\'e} Claude Bernard Lyon 1, CNRS, Institut Lumière Matière, UMR5306, F69100 Villeurbanne, France}

\makeatletter
\AtBeginDocument{\let\LS@rot\@undefined}
\makeatother

\begin{document}

\title{Molecular insight on ultra-confined ionic transport in wetting films: the key role of friction}

\author{Aymeric Allemand}
\altaffiliation{Current address: TIPs, Université libre de Bruxelles, Avenue F.D. Roosevelt 50, Brussels, Belgium}
\affiliation{\ilm}
\author{Anne-Laure Biance}
\affiliation{\ilm}
\author{Christophe Ybert}
\affiliation{\ilm}
\author{Laurent Joly}
\email{laurent.joly@univ-lyon1.fr}
\affiliation{\ilm}

\date{\today}

\begin{abstract}
Nanofluidic transport is ubiquitous in natural systems from extra-cellular communication in biology to geological phenomena, and promotes the emergence of new technologies such as energy harvesting and water desalination. While experimental access to ultraconfined fluids has advanced rapidly, their behavior challenges conventional theoretical descriptions based on Poisson-Boltzmann theory or the Stokes equation whose possible extension remains an open question. In this work, we use molecular dynamics simulations to investigate ionic transport within wetting films of water confined on silica surfaces down to the sub-nanometer scale. We then analyze these results using a simple one-dimensional theoretical framework. Remarkably, we show that this model remains valid even at confinement close to the molecular scale. Our results reveal that ion dynamics play a key role in ionic transport, through ion adsorption at the water-silica interface. Adsorbed cations do not participate in ionic conduction, but instead generate molecular-scale roughness and transmit additional frictional forces to the substrate. This mechanism produces an apparent viscosity increase in electrostatically driven flows, reaching up to four times the bulk value in the case of potassium. Our findings highlight the critical role of interfacial ion adsorption in nanoscale hydrodynamics and provide new insights for interpreting experiments and designing nanofluidic systems.
\end{abstract}

\maketitle

\section{Introduction}
\label{sec:intro}

Water nanofluidics---the study of water transport under nanometric confinement---has attracted increasing attention for its broad range of applications \cite{bocquet_nano_2010}. Inspired by natural systems such as transmembrane nanopores \cite{doyle1998structure} and water infiltration in geological processes \cite{McCarthy_1989_SubsurfaceTO, Mehmani_2019}, researchers can now engineer artificial devices for filtration \cite{Deng_2015}, desalination \cite{Deng_2015,Yang_2019_RO}, energy harvesting \cite{skilhagen_2010_blueenergy,siria_2017_new} or even neuromorphic computing \cite{noy2023nanofluidic, emmerich2024nanofluidic, kamsma2024brain}. 
The specificity of transport at the nanoscale originates from both the high surface-to-volume ratio in these systems, which amplifies the role of liquid/surface interactions on ionic and liquid transports, and the modification of the liquid or ion properties under extreme confinement, these two effects being often coupled.
Interestingly, nanoscale transports and their coupling at liquid-solid interfaces, the so-called electrokinetic phenomena, are usually well captured by continuum descriptions when suitable boundary conditions (e.g., friction coefficients or surface potentials) are specified \cite{bocquet_nano_2010}.

However, these standard physical descriptions have been challenged. For instance, flows can modify surface chemistry at interfaces \cite{lis2014liquid}, and in some systems, quantum effects \cite{Kavokine2022} and solid-liquid interactions \cite{bocquet_nano_2010, bocquet2020nanofluidics} yield unusual fluid behaviors, with implications for materials science \cite{radha2016molecular,esfandiar2017size,xu2018nanofluidics}. Extreme confinement can cause bulk properties to change dramatically, such as the dielectric permittivity of water dropping from 80 to 2 for a $1\,\si{\nano\meter}$ film \cite{fumagalli2018anomalously,Jalali_2021_eps}, or the viscosity of aqueous electrolytes increasing severalfold below $3\,\si{\nano\meter}$ \cite{raviv2004fluidity, bonthuis2012unraveling,Hoang2012a}. At these scales, bulk and surface effects intertwine. Recent simulations by Luan \textit{et al.} \cite{Luan_Lu_Xie_2025} showed that counterion dissociation can make streaming conductance pressure-dependent and decrease the system’s friction coefficient when it is predominantly governed by bound ions. Such complex responses inspire new water-based functional systems \cite{bocquet2020nanofluidics}. Understanding these phenomena calls for molecular-level investigations of interfacial transport in sub-nanometric films.

Before tackling complex coupled transports, a key parameter in most nanofluidic applications is the device's ionic conductance, which often limits its efficiency—such as energy conversion yields \cite{van2007power, siria2013giant} or current rectification level \cite{vlassiouk_nanofluidic_2007, karnik_rectification_2007}.
Classically, surface interactions influence ion transport via surface conductivity \cite{hunter2001}. Ionic conductivity thus has two main contributions: a bulk term, set by the concentration of charge carriers (from salt or pH), and an interfacial term, arising from processes like surface ionization that release counterions. At low salt, the latter dominates, and the system enters the surface-conductivity regime \cite{stein2004surface, siria2013giant}, with channel conductance reaching a plateau as salt concentration decreases.

In that context, experimental studies \cite{liu2010translocation, secchi2016scaling,balme2015ionic,mouterde2019molecular,aguilella2014lipid} put into light some discrepancies with this standard picture of the surface-conductance regime, and its expected conductance plateau. 
These deviations could be attributed to complex charge dissociation mechanisms at the interface \cite{ninham1971electrostatic}, interfacial slip and abnormal charge mobility at the interface \cite{mouterde2019molecular, vinogradova2021enhanced}.  
More recently, our group developed an experimental setup to continuously control the confinement reaching sub-nanometric confined systems, with an access allowing conductance measurements \cite{allemand_2023_cond, leivas2024condensationeffecttransportalumina}.
The experimental results were interpreted by incorporating a single-molecule-thick layer near silica with severely restricted transport, beyond which continuum, bulk-like models could be applied. Although these findings were instructive, the origin at the molecular level of this apparent stagnant layer remains non-elucidated.

Another complementary approach is thus needed. Molecular dynamic (MD) simulations is one of such tools to investigate possible molecular effects in nanometric confinement and to build model system. 
MD simulations have indeed been able to investigate transport properties in specific water-based systems \cite{park_2014_CNT, joly2014, majidi_2022_wat}. But they have mainly focused on the equilibrium interactions with surfaces, such as silica \cite{rimola_2013_sio2, chen2019molecular} or other oxides \cite{tuan_2011_sio2, Wang_2021_oxide}, providing detailed insights into electrolyte behavior at interfaces \cite{dewan_2014_watstrcture, hartkamp_2015, dopke_2019_IFF} that complement continuum mean-field theories. The consequences of such equilibrium interactions on evaporative transport in wetting films \cite{montazeri_2020}, on the disjoining pressure in ultrathin water films \cite{hu_2013}, on electric-field effects on liquid-vapor interfaces \cite{nikzad_2017}, have also been explored, and points the role of the chemical nature of the ions, or more globally ion specificity effects. Ion transport in wetting films on silica \cite{Lentz_2020} has also been considered, showing mainly that some ionic specificity exists at interface.

Establishing a rationalizing framework that accounts for such complex local interactions remains however a key challenge. A possible route consists in extending continuum-based approaches through a proper incorporation of modified boundary conditions or cut-off parameters. Yet, the origin of these model inputs, their dependence on the specific ion and ion-surface interactions, as well as the physical mechanisms underlying the reduction in ion dynamics, remain important open questions.
The objective of this work is then to explore at the molecular scale the conductance behavior of extremely confined wetting films. 
By leveraging MD simulations, we first characterize the wetting film structure and show that the surface charge density and roughness indeed favors cation adsorption, depending on ion nature. But, in a second step, when considering transport, it appears that the mechanism at stake is more complex, and we then explore how this static adsorbed layer can induce an extra friction on the ions, which also depends on the ion nature.
These findings allow us to refine a simplified 1D model that captures the essential features of ion transport under strong confinement.

\section{Materials and methods}

\begin{figure}[htbp]
\includegraphics[width=0.9\linewidth]{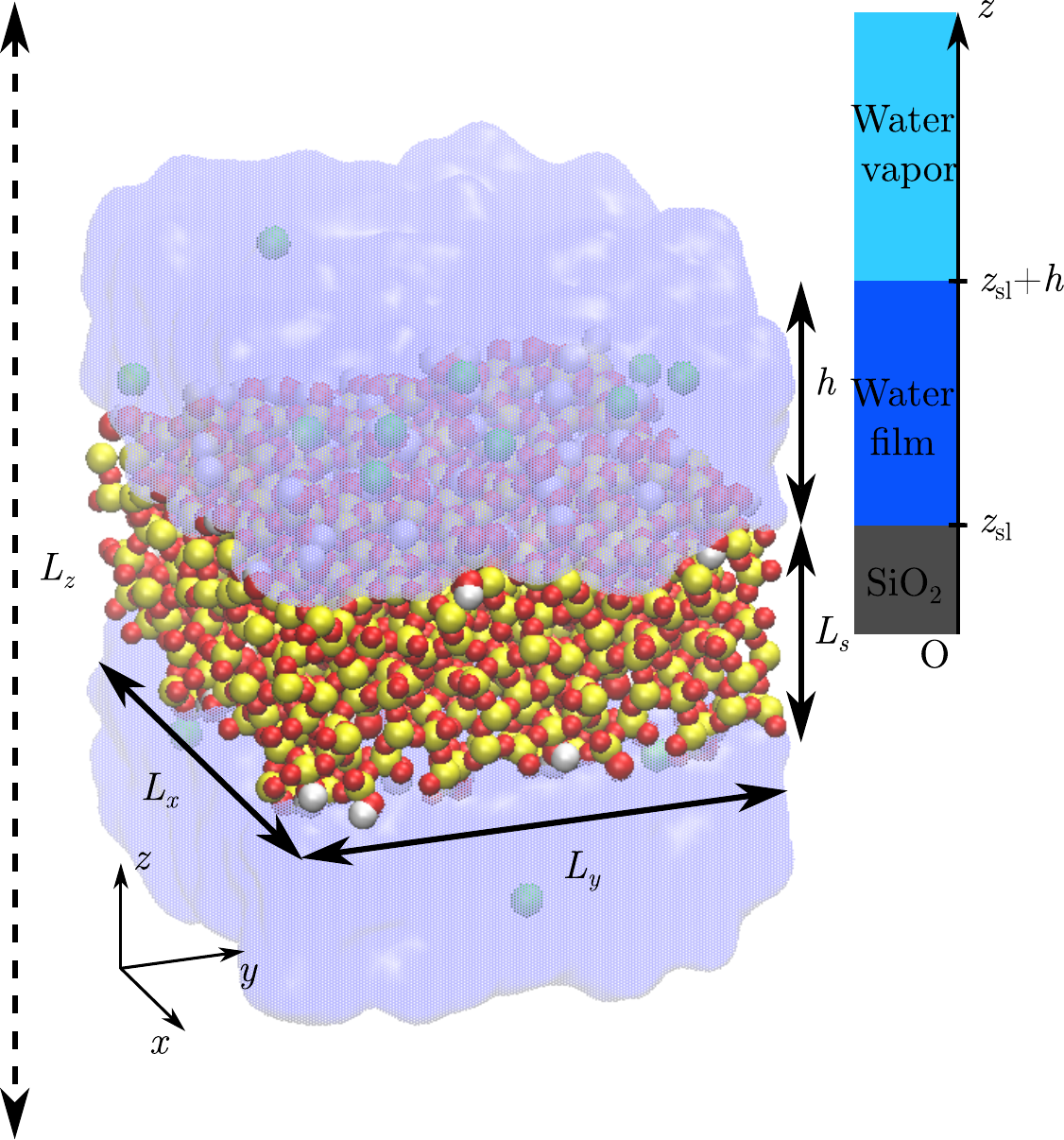}
\caption{Numerical system showing two water films on opposite sides of a charged SiO$_2$ substrate. Periodic boundaries are set in all three dimensions.
Counter-ions are represented as green spheres, Si atoms as yellow, O atoms as red, H atoms as white, and water molecules as a blue translucent envelope. 
The schematic on the right illustrates the simplified one-dimensional system used in the theoretical framework.}
\label{fig:system}
\end{figure}

We performed molecular dynamics (MD) simulations of wetting films adsorbed on a charged silica substrate using the Large-scale Atomic/Molecular Massively Parallel Simulator (LAMMPS) code \cite{thomson_2022_lammps}. The simulated system consists of an amorphous (or crystalline) silica substrate, exposing a density of $\qty{4.5}{SiOH/\nano\meter\squared}$ silanol (SiOH) groups at the surface, consistent with experimental values \cite{zhuravlev2000surface}. On each side of the substrate, a wetting film is present, as shown in Fig.~\ref{fig:system}, whose thickness is set by the number of water molecules, $N_{w}$.
To introduce surface charge, 10 surface silanol groups are randomly ionized on each face of the substrate. To ensure overall charge neutrality, 10 counter-ions (Na$^+$, K$^+$, or Li$^+$) are then added to each wetting film. Each film thus contains $N_{w}$ water molecules and 10 cations. 
Here, using alkali cations provides a practical way to probe ion-specific adsorption and interfacial friction within standard classical MD. 
Note, however, that additional complexity may arise in experimental wetting films on silica, where H$^+$ should act as counter-ions, with special transport mechanisms. We therefore stress that our simulations should be viewed as model systems inspired by silica wetting-film experiments, rather than a fully quantitative description of a specific pH-controlled situation. 

The simulation box measures $L_x \times L_y \times L_z = 4.201 \times 4.195 \times 16.0\,\unit{\nano\meter\cubed}$ for the amorphous silica substrate, and $4.275 \times 4.480 \times 16.0\,\unit{\nano\meter\cubed}$ for the crystalline case. The substrate thickness is $L_s = \qty{2.330}{\nano\meter}$ (amorphous) or $L_s = \qty{2.593}{\nano\meter}$ (crystalline). Periodic boundary conditions are applied in all three directions. The large $L_z$ dimension introduces a vacuum region of at least 10\,nm between periodic images, avoiding artificial interactions along $z$ (one should note also that the simulated systems are electrically neutral, and have no net electric dipole as they are symmetric, which limits the interactions between the periodic images). For one system, we performed simulations with a reduced gap and with an increased gap between the periodic images, and we did not observe any significant difference, see the \textit{supplementary material}. 

The random ionization of surface groups is performed in three independent ways on each face, resulting in six independent wetting film configurations at constant $N_{w}$. Unless stated otherwise, error bars on measured quantities $\mathcal{M}$ represent the standard error over these six realizations, i.e., $\Delta \mathcal{M} = \mathrm{std}(\mathcal{M}) / \sqrt{6}$, where $\mathrm{std}(\ldots)$ denotes the standard deviation.

We used the TIP4P/2005 water model \cite{abascal_2005_tip4p} to simulate water molecules. The SHAKE algorithm \cite{Ryckaert_1977_shake} was used to rigidify 
the water molecules. The TIP4P/2005 model is one of the most accurate \cite{Zaragoza_2019_TIP4P} and has been widely used to study confined systems such as water in carbon nanotubes \cite{Dix_2018_TIP4P,mendonca_2020_TIP4P}.
The cations were described using the Madrid force field, based on the TIP4P/2005 model and using a scaled charge of $0.85$ electron units for ions \cite{Zeron_2019_Madrid, blazquez_2022_Madrid}. This approach provides accurate modeling of solution properties, such as density at high concentrations, viscosity, and solubility. As a validation of the implementation, we computed the bulk diffusion coefficient of Na$^+$, K$^+$, Li$^+$ and H$_2$O for these specific force fields. The obtained values (see table S5 in the \textit{supplementary material}) are in good agreement with previous studies \cite{abascal_2005_tip4p,Zeron_2019_Madrid}. More details can be found in the \textit{supplementary material}. 
Additionally, the Interface Force Field \cite{emami_2014_FF} has been used for the liquid-solid interaction parameters. This force field has been shown to provide a good description of ion adsorption on charged amorphous silica when combined with the same water model as in the present study, namely TIP4P/2005 \cite{dopke_2019_IFF}. Compared to classical force fields used to model silica, such as ClayFF, which show good agreement with experimental bulk properties \cite{Mishra2017}, the Interface Force Field was specifically designed and benchmarked to reproduce interfacial properties of silica in contact with water, such as water adsorption isotherms and water contact angles \cite{emami_2014_FF}.
Lastly, the cutoff parameter for the interaction potential between atoms is set to $\qty{10.0}{\angstrom}$ and long range coulomb interactions are computed thanks to the particle-particle/particle mesh (PPPM) method \cite{Hockney_1988_PPPM} with a relative error in forces set to $10^{-6}$.

The timestep is set to $\delta t = \qty{1}{\femto\second}$. 
A Nose--Hoover chain thermostat with a chain length of three and 
a damping time of $\qty{100}{\delta t}$ is used to set a constant temperature $T_0 = \qty{300}{\kelvin}$. 
The {equations} of motion are applied to water molecules, cations and to the oxygen and hydrogen atoms of silanol groups. A typical simulation is as follows: $2N_{w}$ water molecules and $2 \times 10$ cations are placed on the silica substrate (on two opposite sides) and a step of energy minimization occurs to avoid particle superposition. The system is equilibrated for $\qty{0.5}{\nano\second}$. Equilibrium measurements are taken during the next $\qty{1.0}{\nano\second}$ 
with no applied external forces. 
For conductance measurements, an external electric field $E_x$ is applied to the fluid along the $x$ direction. The first $\qty{0.1}{\nano\second}$ are used to reach a steady state, and the subsequent $\qty{1.0}{\nano\second}$ are dedicated to measuring the ionic current while monitoring temperature, pressure, and energy to ensure their stability. The applied electric field values range from  $E_x = \qty{4.25}{\milli\volt\per\angstrom}$ to $E_x = \qty{34.0}{\milli\volt\per\angstrom}$, for which we verified that the system remained in the linear response regime.

{To measure the thickness $h$ of the wetting film formed onto the silica interface, we use the following equation:}
\begin{equation}
    \rho_\textrm{b} h L_x L_y = N_\textrm w m_{\textrm{H}_2 \textrm{O}}
    \label{eq:Nw_h}
\end{equation}
{with $L_i$ the length of the simulation box in the $i$-direction ($i=x,y$), $\rho_\textrm{b}$ is the water bulk density \cite{abascal_2005_tip4p} and $m_{\textrm{H}_2 \textrm{O}}$ is the mass of a water molecule. Equation \eqref{eq:Nw_h} is a consequence of the Gibbs dividing plane method \cite{botan_2013_GDP, Lamorgese_2017_GDP, herrero_2019_GDP} given, in our case, by the following equality:}
\begin{equation}
    \int_0^{+\infty} \rho(z) \dd z = \int_{z_\textrm{sl}}^{z_\textrm{sl}+h} \rho_\textrm{b} \dd z
    \label{eq:int_Nw_h}
\end{equation}
{with $\rho(z)$ the density profile of the MD simulation and $z_\textrm{sl}$ the position separating a region of homogeneous fluid and a region of solid. This integrative form has been used to determine the position $z_\textrm{sl}$ that has been used as a fitting parameter for simulations with $h>\qty{1}{\nano\meter}$. It appears that the position $z_\textrm{sl}$, shown in table \ref{tab:fit_param}, does not depend on 
the side of the SiO$_2$ substrate on which it forms}, the type of cations and the wetting film thickness.
The reader is referred to the \textit{supplementary material} for additional numerical details.

\section{Results and discussion}

\begin{figure}[htbp]
    \centering
    \includegraphics[width=0.9\linewidth]{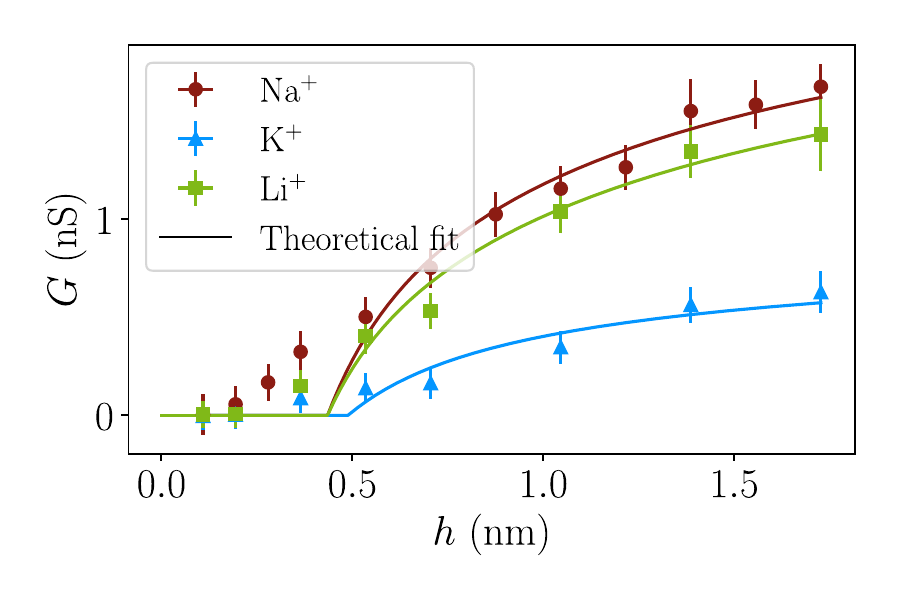}
    \caption{Conductance $G$ as a function of the thickness of the water film $h$ for three different cations (Na$^+$, K$^+$ and Li$^+$). Solid lines correspond to theoretical fits from a 
    {simple continuum} model {proposed} in Ref.~\cite{allemand_2023_cond}.}
    \label{fig:G_vs_h_NumTh}
\end{figure}

We start by computing the conductance of the wetting films, 
{that we define} as the ratio between the (ionic) electric current $I$ in the film and the applied voltage $\Delta V$: 
\begin{equation}
    G = \frac{I}{\Delta V} = \frac{I}{E_x L_x}.
\end{equation}
To measure the wetting film's conductance, we apply a series of increasing electric fields, sum the resulting ionic currents from all cations, and determine the conductance from a linear fit to the current-voltage curve. 
Figure~\ref{fig:G_vs_h_NumTh} shows the numerical conductance $G$ 
as a function of the water film thickness $h$. The data include results for the three cations, Na$^+$, K$^+$, and Li$^+$, on an amorphous silica substrate.
For all cations, the {trend is similar: the conductance vanishes at a finite film thickness, above which it increases sublinearly with $h$.}
{However, the amplitude of the conductance for K$^+$ is lower than for Na$^+$ and Li$^+$, which exhibit very comparable values.}

The observed increase in $G$ with $h$ stands in contrast to the expected behavior in the surface-conductivity regime. In principle, the number of charge carriers within the liquid film is governed solely by the surface charge and is therefore independent of $h$, implying that $G$ should remain unaffected by variations in $h$~\cite{stein2004surface, duan2010anomalous, siria2013giant, kavokine2021fluids}.
{However, these numerical results are in full agreement with experimental measurements performed on a water-vapor film condensed onto silica surfaces \cite{allemand_2023_cond}, which demonstrates that our numerical system successfully captures the phenomenology of ultra-confined systems.}

{In the experimental study,} 
{we} {proposed that this failure of the surface-conductivity regime arises due to the presence of a \emph{hindered mobility layer} for charge carriers near the surface. Indeed, combining a continuous classical approach based on the Poisson--Boltzmann (PB) theory~\cite{bocquet_nano_2010, gravelle_nanofluidics_2016, herrero2021poisson, herrero2022chapter} to describe the distribution of cations with such a motionless layer accounted for experimental data.}
In Fig.~\ref{fig:G_vs_h_NumTh}, the solid curves correspond to fits obtained from this model, using two adjustable parameters: the hindered-layer 
thickness $\delta_\text{hdl}$ and the constant surface-charge density $\Sigma_\text{fit}$. The best-fit values of these parameters are summarized in Table~\ref{tab:fit_param}.

\begin{table}[b]
    \caption{Fitting parameters for the continuous model developed in reference \cite{allemand_2023_cond} for the numerical conductance for the three different cations: Na$^+$, K$^+$, and Li$^+$ and position of the solid-liquid $z_\textrm{sl}$ determined by equation \eqref{eq:int_Nw_h}.}
    \begin{ruledtabular}
        \begin{tabular}{ccc|c}
        Cation &
        $\Sigma_\text{fit}/\Sigma$ (\si{\percent})& $\delta_\text{hdl}$ (\si{\nano\meter})& $\lvert z_\textrm{sl} \rvert$ (\si{\nano\meter})\\
        \colrule
        Na$^+$ & $68.8\pm 1.1$ & $0.435 \pm 0.043$ & $0.96 \pm 0.02$ \\
        K$^+$ & $13.9 \pm 0.5$ & $0.489 \pm 0.011$ & $0.98 \pm 0.02$\\
        Li$^+$ & $84.4 \pm 1.4$ & $0.435 \pm 0.014$ & $0.98 \pm 0.02$\\
        \end{tabular}
    \end{ruledtabular}
    \label{tab:fit_param}
\end{table}

{Qualitatively, this simple description performs well, and we can now examine its associated parameters. As in experiments~\cite{allemand_2023_cond}, the}
hindered-layer 
thickness, $\delta_\text{hdl}$, is 
close to the diameter of a water molecule. 
This finding is in line 
with the concept of a structured stagnant water layer~\cite{georges_drainage_1993,cieplak2001boundary}. Moreover, $\delta_\text{hdl}$ appears to be independent of the cation and thus of their size \footnote{Indeed, the {Van der Waals} diameters of the cations are $2.22\,\si{\angstrom}$ for Na$^+$, $2.30\,\si{\angstrom}$ for K$^+$, and $1.44\,\si{\angstrom}$ for Li$^+$ (obtained from {the equilibrium distance of the ion-ion Lennard-Jones potential}).}.   
However, when considering the second fitting parameter, the surface-charge density, $\Sigma_\text{fit}$, it differs from the actual set value of $\Sigma \approx 85\,\text{mC/m}^2$, which corresponds to the ten ionized silanol groups at the surface. 
Specifically, $\Sigma_\text{fit}$ amounts to about $68.8\,\%$ and $84.4\,\%$ of $\Sigma$ for Na$^+$ and Li$^+$, respectively, but only $12.7\,\%$ of $\Sigma$ for K$^+$. 
{This quantitative mismatch suggests the underlying molecular mechanisms are likely more intricate than suggested by a simple hindered diffusion model.}

To examine these mechanisms in greater detail, we first recall the expression of the conductance (within a one-dimensional model):
\begin{equation}
    G = \frac{L_y}{L_x} \left\{ \int e^2 c_+(z) \mu_+(z) \, \mathrm{d}z + \int e c_+(z) \frac{v_x(z)}{E_x} \, \mathrm{d}z \right\},
    \label{eq:G_contrib}
\end{equation}
where the integrals span over the whole system. In this equation, the distribution of cations along the confinement direction $z$ is represented by $c_+(z)$, while $\mu_+(z) = D_+(z)/(k_\mathrm{B} T)$ is their local mobility {as determined from} their diffusion coefficient $D_+(z)$.
Finally, $v_x(z)/E_x$ characterizes the electro-osmotic velocity profile of the solvent normalized by the applied electric field.
Physically, the first integral in Eq.~\eqref{eq:G_contrib} corresponds to the electrophoretic contribution, arising directly from the motion of cations under the applied electric field. The second integral describes the electro-osmotic contribution, resulting from the advective transport of cations carried along by the solvent's electro-osmotic flow.

In the following, we will use MD simulations to investigate separately the profiles of cation density $c_+(z)$, fluid velocity $v_x(z)$ and cation mobility $\mu_+(z)$. {This will eventually provide a detailed description of the different contributions to ion transport in ultra-confined liquid layers.}

\subsection{Ion distribution}

\begin{figure}[htbp]
    \centering
    \includegraphics[width=0.9\linewidth]{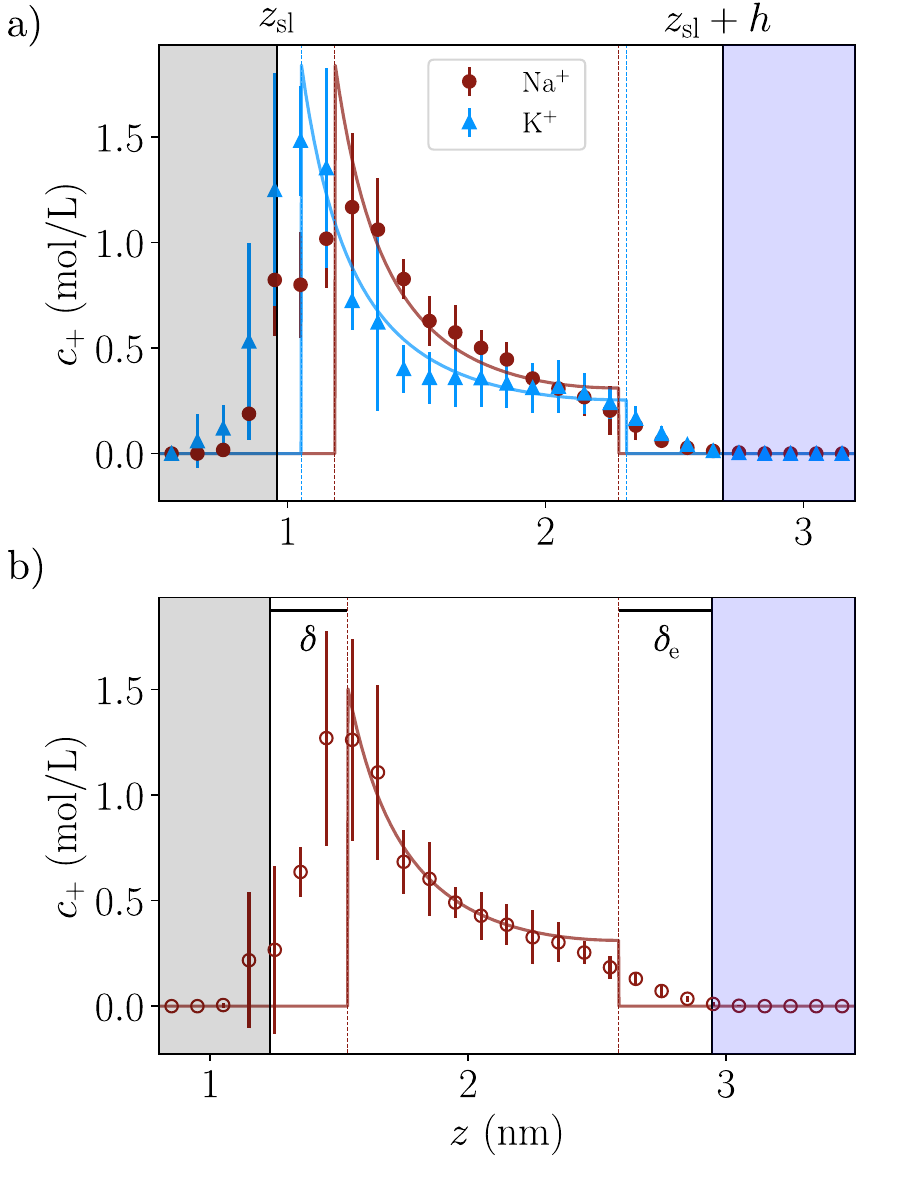}
    \caption{Typical cation concentration profiles as a function of the vertical position $z$ in the wetting film, for Na$^+$ (circles) and K$^+$ (triangles) cations with $h = \qty{1.73}{\nano\meter}$.
    a) amorphous SiO$_2$; b) crystalline SiO$_2$. 
    $z_\mathrm{sl}$ {locates the solid-liquid interface } 
    and $h$ is the wetting film thickness {(see Materials and Methods)}.
    Two layers are shown of thickness $\delta$ and $\delta_\mathrm e$, which are discussed in the main text. The colored solid lines correspond to an electrostatic model further discussed in the main text.
    }
    \label{fig:cp_vs_z_hmax}
\end{figure}

First, we consider the cation distribution as a function of the vertical position $z$ in the wetting film. 
Typical cation concentration profiles $c_+(z)$  are presented in Fig.~\ref{fig:cp_vs_z_hmax} {for a film thickness $h = \qty{1.73}{\nano\meter}$ on amorphous (a) or crystalline (b) SiO$_2$ substrates.} 
Additional profiles for various thicknesses and cations are provided in the \textit{supplementary material} (see Figs. S4, S5, and S6).
{We recall that interfaces are set by the Gibbs dividing plane method \cite{herrero_2019_GDP}, with $z=z_\mathrm{sl}$ and $z=z_\mathrm{sl}+h$ the solid-liquid (respectively, liquid-vapor) interfaces (see also the inset of Fig.~\ref{fig:system}).}

The concentration profiles behave as expected in the central region of the wetting film, decaying with the distance to the wall.
However, departures from this simple picture {are observed near both film interfaces.} 
First, at the solid-liquid interface, the position $z_\mathrm m$ of the concentration maximum is located above the Gibbs dividing surface $z=z_\mathrm{sl}$. This is reminiscent of MD simulations performed near silica in the presence of salt \cite{chen2019molecular} and can be explained by the molecular roughness of the surface, specifically by the protuberance of silanol groups within the wetting films. In the following, we will denote $\delta = z_\mathrm m - z_\mathrm{sl}$. 

\begin{figure*}[htbp]
    \centering
    \includegraphics[width=1.0\textwidth]{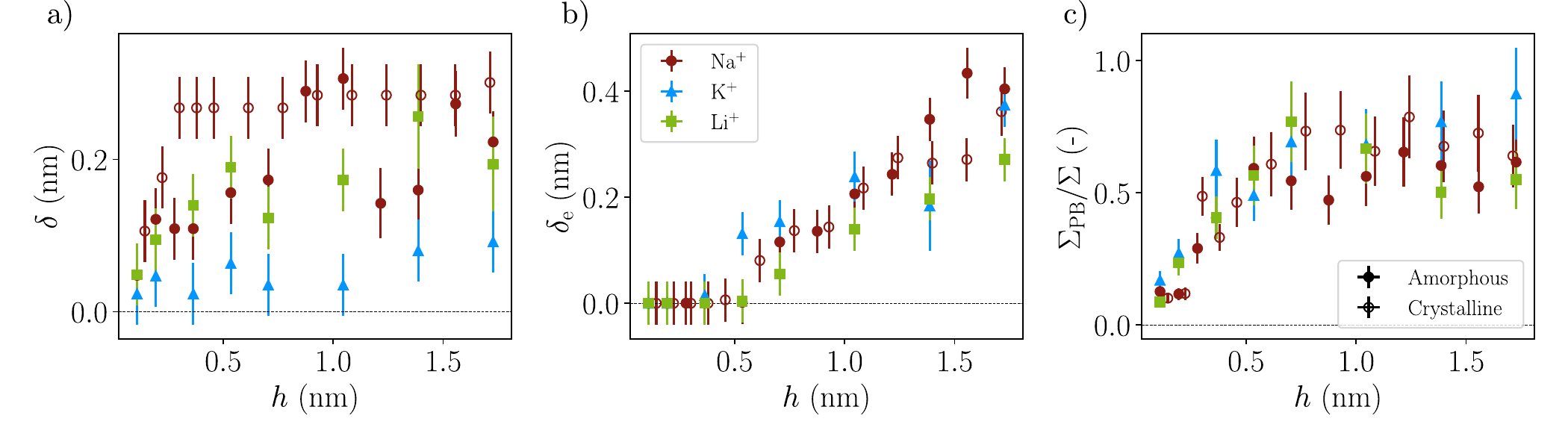}
    \caption{{Evolution of the ion} {profiles parameters with the film thickness.}
    (a) {Shift between solid-liquid and surface charge planes $\delta$;} 
    (b) {Image charge repulsion layer $\delta_\mathrm e$;}
    and (c) Effective surface charge density $\Sigma_\mathrm{PB}$ 
    normalized by the {nominal} surface charge $\Sigma$.} 
    \label{fig:d_de_S_vs_h}
\end{figure*}

Second, ions are depleted from the liquid-vapor interface. 
This is due to image charge repulsion {resulting from}
the jump of dielectric permittivity between liquid and vapor \cite{Samaras_1934,huang_2007, huang_aqueous_2008, LeBreton_2020}. 
In this work, we compute the thickness of the cation exclusion layer $\delta_\mathrm e$ by defining a concentration threshold $\delta c_\mathrm{th} = \qty{0.15}{\mole\per\liter}$ (the choice of the value of $\delta c_\mathrm{th}$ is detailed in the \textit{supplementary material}).
For each ion type, substrate type, and film height, we use the data from six independent ion density profiles to determine both $\delta$ and $\delta_e$.

{Once these surface effects are considered,}
i.e. for $z \in [z_\mathrm{sl} + \delta$ ; $z_\mathrm{sl} + h - \delta_e]$, 
{density profiles are well described
by continuous PB theory, despite the strong confinement
regime:}
\begin{equation}\label{eq:PBprofile}
    c_+(z) = \dfrac{K^2}{2 \pi \ell_\mathrm{B}} \exp \left( \dfrac{-eV(z)}{\kB T} \right) , 
\end{equation}
with the electric potential $V(z)$ expressed as:
\begin{equation}
    V(z) =  \frac{\kB T}{e} \log \left( \cos^2 \left( K (z-h-\delta_\mathrm e)\right) \right) ,\\
    \label{eq:V_z}
\end{equation}
where $\ell_\mathrm{B}$ is the Bjerrum length, $e$ the elementary charge, $\kB$ the Boltzmann constant, $T$ the temperature, and $K$ is the inverse of a length that is linked to the Gouy-Chapmann length $\ell_\mathrm{GC}$ (and hence to the effective surface charge density $|\Sigma_\mathrm{PB}| = e/(2\pi \ell_\mathrm{B} \ell_\mathrm{GC})$) through: 
\begin{equation}
        K (h - \delta - \delta_\mathrm e) \tan \left( K (h - \delta - \delta_\mathrm e) \right) = \frac{h - \delta - \delta_\mathrm e}{\ell_\mathrm{GC}} .
    \label{eq:K_lGC}
\end{equation}
Indeed, Fig.~\ref{fig:cp_vs_z_hmax} shows that Eq.~\eqref{eq:PBprofile} describes well the numerical results $c_+(z)$ in the central region, regardless the crystalline or amorphous state of the SiO$_2$ substrate.
{From the associated values for $K$, we can recover }
the effective surface charge $\Sigma_\mathrm{PB}$.

The evolution of the parameters $\delta$, $\delta_e$, and $\Sigma_\mathrm{PB}$ with the film height $h$ for a variety of systems is shown in Fig.~\ref{fig:d_de_S_vs_h}.  
Let us first focus on the shift $\delta$ between the location of the maximum ion concentration and that of the solid-liquid interface. For Na$^+$ ions and on crystalline silica, $\delta$ is constant at a value of $\sim 0.3\,\si{\nano\meter}$ except for the two thinnest films where the thickness goes below the molecular size. This is consistent with the idea that the shift mostly arises from the protrusion of the charged surface groups inside the film. The shifts $\delta$ are more dispersed in the case of the amorphous substrate, due to the additional roughness of this substrate, but also reach $\sim 0.3\,\si{\nano\meter}$ for the thickest films, and take lower values for the thinnest ones. 

Overall the values of $\delta$ are lower on the amorphous substrate. Indeed, one could argue that on the rougher amorphous substrates, the shift between the Gibbs dividing surface and the average position of the charged surface groups could be reduced, and that the counter-ions should have more freedom to align with the effective position of the wall surface charge. Finally, one should note that $\delta$ takes similar values for Na$^+$ and Li$^+$, but that the values are smaller for K$^+$. 
We will come back to that result later, when discussing the differences in ion adsorption.

At the liquid-vapor interface, the situation is quite different, as $\delta_e$ is the same for the 3 ions, and decays from around one molecular size in the thickest films, to zero for film thicknesses below $0.5\,\si{\nano\meter}$. The absence of ion specificity tends to indicate that the exclusion at the liquid-vapor interface is indeed controlled by electrostatics. 

{
Finally, Fig.~\ref{fig:d_de_S_vs_h}(c) shows the evolution of the effective surface charge $\Sigma_\mathrm{PB}$ with the film thickness.
As we can see, it is mostly independent of the ion or SiO$_2$ substrate and displays an increase with film thickness until it plateaus beyond $h > 0.6\,\si{\nano\meter}$ at values from \qtyrange{60}{80}{\%} of the nominal surface charge. Such value is consistent with the fact that part of the counter-ions lie in the molecular roughness region. 
Also striking is the fact that the different ions  
}
show similar static behavior, at odd with the drastically different conductance measured for K$^+$. This points to a critical role of hydrodynamics, which we will explore in the next section.

%
%
\subsection{Dynamics}
%
%
%
Having characterized the static distribution of the ions in the wetting film, we now turn to the dynamical properties of the film. 
Charge carriers have two contributions to the conductance: through the advection of the charge by the surrounding flow (electro-osmosis) and through electrophoresis, i.e. the motion of a charged particle, relatively to the solvent, under an electric field. We will tackle these two aspects in this section.

\subsubsection{Electro-osmotic velocity}

\begin{figure}[htbp]
    \centering
    \includegraphics[width=0.9\linewidth]{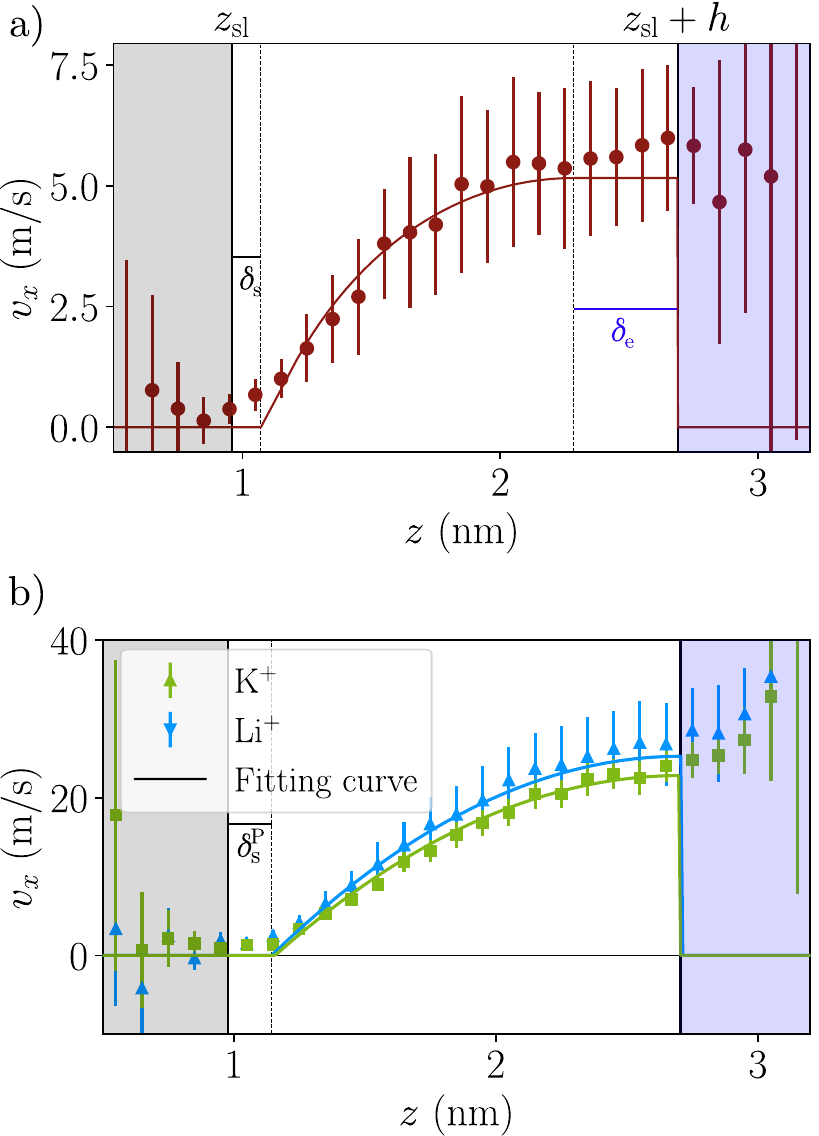}
    \caption{(a) Typical electro-osmotic velocity profile $v_x(z)$ for Na$^+$ at $E_x = \qty{34}{\milli\volt\per\angstrom}$ on an amorphous SiO$_2$ substrate. The colored solid line corresponds to the theoretical profile. (b) Typical pressure induced velocity profile $v_x(z)$ for K$^+$ (light blue data) and Li$^+$ (green data) for an applied force {per particle} $|f| = \qty{8.0}{\calorie\per\angstrom\per\mole}$ on an amorphous SiO$_2$ substrate. The velocity profile for Na$^+$ is similar to the Li$^+$ one. These profiles are extracted from a water film of thickness $h = \qty{1.7}{\nano\meter}$. The colored solid lines correspond to hydrodynamic model curves (see main text). In both graphs, the black solid vertical lines correspond to the positions $z=z_\text{sl}$ and $z=z_\text{sl}+h$. The dotted black lines correspond to the position of the no-shear plane $z=z_\text{sl}+\delta_\text{s}^i$, with $i=\varnothing,$P (more details in the text below) and the dotted blue vertical line is the position $z=z_\text{sl}+h-\delta_\text{e}$. 
    }
    \label{fig:v_vp_vs_z_hmax} 
\end{figure}

The applied electric field induces a coulombic force on the non-neutral part of the fluid
{that generates a so-called }
electro-osmotic flow \cite{bocquet_nano_2010}. Figure~\ref{fig:v_vp_vs_z_hmax}(a) displays a typical velocity profile for an applied electric field $E_x = \qty{34}{\milli\volt\per\angstrom} $ for Na$^+$ on an amorphous SiO$_2$ substrate {(see the \textit{supplementary material} Figs.~S7, S8 and S9 for additional profiles with different ions and film thicknesses).}
{As expected, the velocity increases from zero nearby the wall up to a maximum at the liquid-vapor interface.
Note that velocity profile can be traced also in the vapor phase, although with very poor statistics; from now on we restrict our discussion to the liquid film.}

{More quantitatively, such a flow is classically described within continuum hydrodynamics framework. At this scale though, it requires proper handling of the boundary conditions.}
Due to substrate roughness and wettability, a stagnant layer of liquid appears \cite{herrero_2019_GDP}, i.e., the flow velocity vanishes slightly above the wall position $ z=z_\mathrm{sl} $. In the following, we will denote by $\delta_\mathrm{s}$ the thickness of the stagnant layer. 
{Stokes equation can then be complemented}
{with the electric force density computed based on the theoretical PB charge density profiles obtained in the previous section. In particular, the theoretical charge density profiles vanish below $z_\mathrm{sl}+\delta$, as illustrated in Fig.~\ref{fig:cp_vs_z_hmax}(a); this is equivalent to considering that ions below $z_\mathrm{sl}+\delta$ are bound to the wall and do not transmit the electric force to the solvent.} More details on the model description can be found in the \textit{supplementary material}. 

\begin{figure}[htbp]
    \centering
    \includegraphics[width=0.9\linewidth]{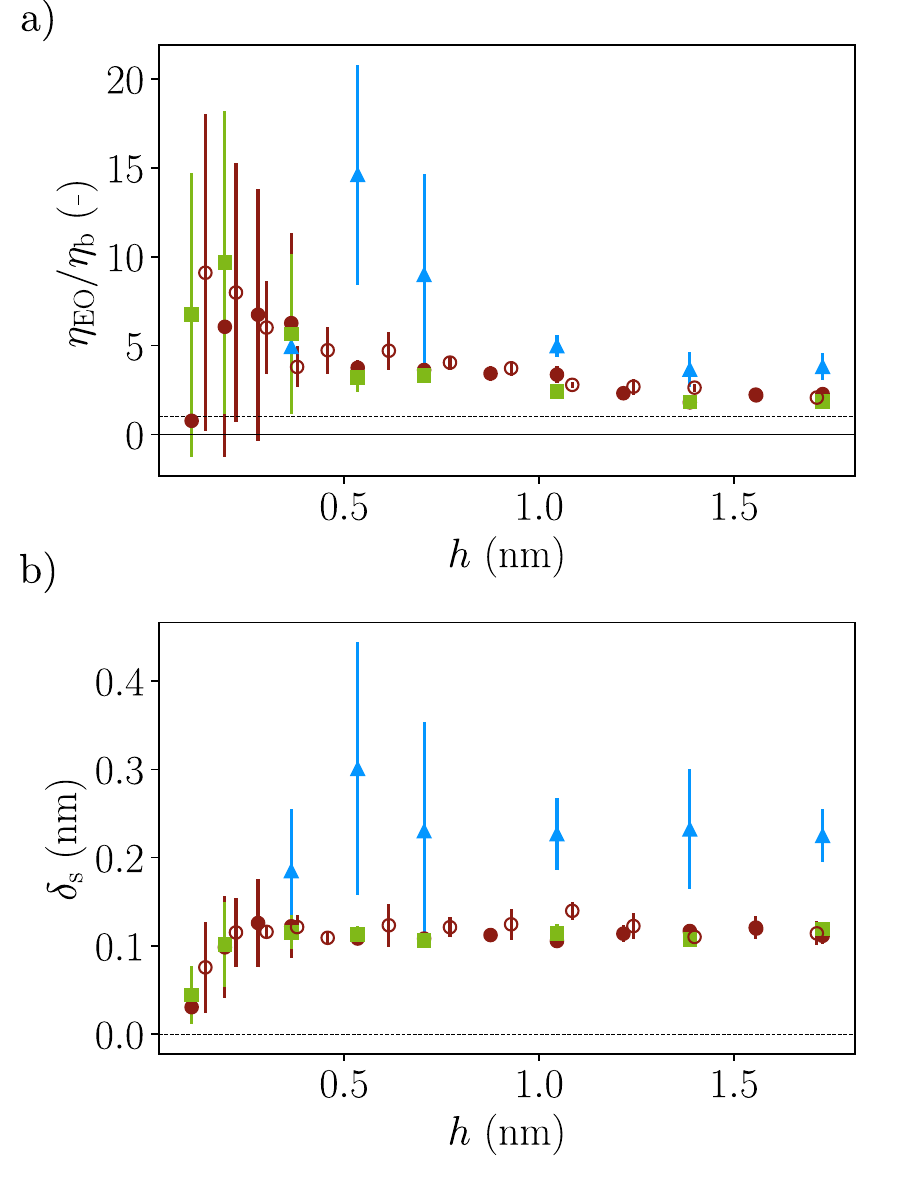}
    \caption{Evolution of the electro-osmotic velocity profiles parameters with the film thickness: (a) effective viscosity $\eta_\text{EO}$, normalized by the bulk value $\eta_\mathrm{b}$ (inset: a zoom in the graph for $h>\SI{1}{\nano\meter}$), and (b) stagnant layer thickness $\delta_\mathrm{s}$.} 
    \label{fig:eta_ds_vs_h}
\end{figure}

Figure~\ref{fig:v_vp_vs_z_hmax}(a) shows that the theoretical velocity profiles obtained with this approach are in good agreement with the numerical results. However, when using the nominal viscosity of the TIP4P/2005 water model ($\eta_\mathrm{b} = 0.83\,\si{\milli\pascal\second}$ at our simulation temperature extracted from reference \cite{MonterodeHijes_2019}), the predicted electro-osmotic flow is significantly overestimated (not shown). To achieve agreement between theory and simulation, it is necessary to introduce an effective viscosity, $\eta_\mathrm{EO}$, for the electro-osmotic flow, which is substantially higher than the nominal bulk viscosity of the water model for the entire range of film thicknesses considered. Note that $\eta_\mathrm{EO}$ is introduced as a global (film-averaged) parameter. Using this homogeneous $\eta_\mathrm{EO}$ is sufficient to reproduce the simulated velocity profiles over the explored range of film thicknesses. We note, however, that the fits are primarily sensitive to the near-wall region where the shear rate is largest; therefore, a possible variation of the viscosity away from the wall would be only weakly constrained by the velocity profiles. 

The variation of this effective viscosity and of $\delta_\mathrm{s}$ with the film thickness $h$ are shown in Fig.~\ref{fig:eta_ds_vs_h}(a) and (b).
The effective viscosity $\eta_\text{EO}$ decreases with increasing $h$ and is expected to reach $\eta_\mathrm{b}$ for $h \rightarrow \infty$; however, it remains significantly larger than $\eta_\mathrm{b}$ for the maximum $h$ that we simulated. For extremely thin films ($h<0.5$\,nm for Na$^+$ and Li$^+$, $h<1$\,nm for K$^+$), the deviations from  $\eta_\mathrm{b}$ become very large, up to a factor of 15 for K$^+$. 
The observed increase in effective viscosity is consistent with the stronger confinement and enhanced ion-substrate friction caused by the protrusion of charged groups at the solid-liquid interface. Notably, the values obtained are equivalent for both amorphous and crystalline substrates, and they are significantly larger for K$^+$ ions than for Na$^+$ and Li$^+$. 
Moreover, the thickness of the stagnant layer is constant for $h>\qty{0.25}{\nano\meter}$ (i.e., comparable to the molecular size) and remains of the same order as $\delta$ for Na$^+$ and Li$^+$. In contrast, the stagnant layer for K$^+$ is approximately twice that of the other two cations. 
This observation is consistent with the stronger frictional forces acting on substrate with absorbed K$^+$. Lastly, the uncertainties associated with the parameters $\eta$ and $\delta_\mathrm{s}$ increase as the water film thickness $h$ decreases. This trend arises because fewer statistical samples are available at smaller wetting film thicknesses. The effect becomes particularly pronounced for $h<\qty{0.5}{\nano\meter}$ in the case of Na$^+$ and Li$^+$, and for $h<\qty{1}{\nano\meter}$ for K$^+$. Moreover, the uncertainties for K$^+$ are larger than for the other two cations, owing to the higher number of K$^+$ ions adsorbed at the interface, which reduces the statistical accuracy of both the cation concentration and electro-osmotic velocity profiles.

\begin{table}
\caption{\label{tab:veo_fit_param}Parameters for the theoretical velocity profiles for Na$^+$, K$^+$, and Li$^+$ on amorphous SiO$_2$ from the analysis of electro-osmotic or pressure driven velocity profiles. These values are averaged values for film thicknesses $h>1\,\si{\nano\meter}$, accordingly with the hydrodynamic continuum limit.}
\begin{ruledtabular}
\begin{tabular}{cccc}
    & Na$^+$ & K$^+$  & Li$^+$ \\ \hline
\multicolumn{4}{c}{Electro-osmotic flow} \\ \hline
$\delta_\mathrm{s} ~$(\si{\nano\meter})  & $0.11 \pm 0.02$ & $0.23 \pm 0.05$ & $0.11 \pm 0.01$ \\
$\eta_\text{EO}/\eta_\mathrm{b}$ (-) & $2.39 \pm 0.52$ & $4.14 \pm 0.80$ & $2.03 \pm 0.27$ \\ \hline
\multicolumn{4}{c}{Poiseuille flow} \\ \hline
$\delta_\mathrm{s}^\mathrm{P} ~$(\si{\nano\meter})  & $0.19 \pm 0.02$ & $0.17 \pm 0.02$ & $0.18 \pm 0.02$ \\
$\eta_\mathrm{P}/\eta_\mathrm{b} ~$(-)  & $1.17 \pm 0.03$ & $1.08 \pm 0.03$ & $1.17 \pm 0.03$
\end{tabular}
\end{ruledtabular}
\end{table}

To investigate the origin of this increase in viscosity, we simulated Poiseuille flows for $h = \qty{1.7}{\nano\meter}$. 
Numerically, an additional force was applied to water molecules to simulate the applied pressure gradient. Additional numerical details are provided in the \textit{supplementary material}. Figure~\ref{fig:v_vp_vs_z_hmax}(b) shows two typical Poiseuille velocity profiles for K$^+$ and Li$^+$ cations, {together with theoretical expectations}. We used this parabolic Hagen-Poiseuille velocity profile to model the numerical data. We considered a stress-free vapor/liquid interface (i.e. at $z=z_\textrm{sl}+h$) and a stagnant layer $\delta_s^P$. 
{As can be seen in table~\ref{tab:veo_fit_param}, this leads to consistent estimations for the stagnant layer thickness.
For the viscosity however, while $\eta_\mathrm P$ indeed shows an increase as compared to $\eta_\text{b}$, it remains of modest amplitude,}
with $\eta_\mathrm P \simeq 1.1 \,\eta_\mathrm b$ for the three ions. 

\label{sec:eo_vel}

This result points toward the idea that the enhanced effective viscosity {evidenced in the electro-osmotic flow} is due to an increased friction between the ions and the substrate, while the intrinsic viscosity of the solvent is not significantly affected by the confinement. In that case indeed, the electric force applied to the ions will be partly transmitted to the substrate, and only a fraction of the force will be transmitted to the solvent and generate the electro-osmotic flow. The effect will be less marked for Poiseuille flows as the external force is transmitted directly to the solvent. We will come back to this idea when discussing the diffusivity of the solvent and of the ions in the following. 

\begin{figure}[htbp]
    \centering
    \includegraphics[width=0.9\linewidth]{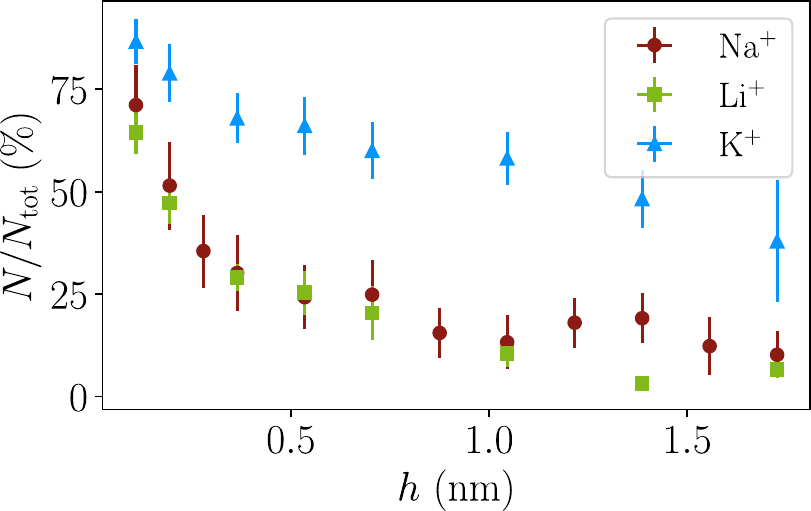}
    \caption{{Time-averaged fraction of ions bound to a charged site of the substrate.}  
    }
    \label{fig:N_vs_z_hmax}
\end{figure}

One possible interpretation for the additional friction experienced by the ions at the substrate interface is that a fraction of ions intermittently adsorb onto charged sites of the substrate. While adsorbed, these ions directly transfer the electrical force acting on them to the substrate. Consequently, averaged over time, a fraction of the electric force is effectively transmitted to the substrate, resulting in an additional frictional contribution from the ions. 
We computed the average number of adsorbed cations based on a static criterion, namely, 
we counted the number of cations that were located in a shell defined by a distance threshold {based on the radial distribution function of each ion around silanol groups (see the \textit{supplementary material} for details).}
The time averaged percentage of cation that are located in the shell of first neighbor (further mentioned as adsorbed cations) is shown on Fig.~\ref{fig:N_vs_z_hmax}. One can see that, for $h>\qty{1}{\nano\meter}$, $10~\%$ of the number of Na$^+$ or Li$^+$ are adsorbed at the interface, this number goes up to $50~\%$ of cations when one considers K$^+$. We can then assume that because of the closeness of cations to negatively charge silanol groups make them strongly coupled, even under a strong electric field. These results are consistent with the larger effective viscosity for EO flows and a smaller $\delta$ value for K$^+$ ions. Indeed, as K$^+$ cations are more bounded, the ions are closer to the substrate thus created a more compact adsorbed ion layer.

A possible interpretation for the stronger adsorption of K$^+$ is related to its hydration properties. Our radial distribution functions (see the \textit{supplementary material}) indicate that, for our force field, K$^+$ is the least strongly hydrated among the three cations (i.e., it has a looser hydration shell). This can reduce the penalty for partial dehydration close to the surface, and therefore facilitate a closer approach and more frequent adsorption events. In this picture, K$^+$ spends more time in the near-wall region where short-range ion--surface interactions are important, which can enhance the effective ion--substrate friction and amplify its impact on the electro-osmotic response. We emphasize, however, that this trend may be specific to the present force field and silica surface model, and should therefore be taken with caution when comparing to experiments. 

Indeed, the adsorption of ions at the silica interface is a complex process that is influenced by interfacial hydrophilicity \cite{calero_2011}, the density of surface silanol groups \cite{videla_2011}, steric effects and ion size \cite{hocine_2016,dopke_2019_IFF}, as well as hydration properties \cite{hartkamp_2015,dopke_2019_IFF}. Despite extensive studies, the adsorption behavior of specific ions remains a matter of debate: some works report stronger binding of K$^+$ to the interface \cite{Morag_2013_ad}, whereas others suggest the opposite trend \cite{hartkamp_2015}. Importantly, simulation outcomes are strongly dependent on the employed molecular dynamics model, and particularly the Lennard-Jones cross-interaction parameters between cation and the surface \cite{hocine_2016}. Therefore, discrepancies observed between cations in our simulations should be interpreted with caution. However, having an enhanced effective viscosity in electrically induced flow for the three ions that are considered remain a good validation of an additional friction force on the surface by the ions that can be included in a simple one-dimensional model.

\subsubsection{Cation mobility}

Apart from the electro-osmotic contribution to the conductance, the cation mobility $\mu_+$ is an important quantity that can be influenced by the hydrodynamic properties of the system, e.g. solvent viscosity \cite{einstein_1905_brown, Miller_1924_Dcoef} or friction at interfaces \cite{Saugey_2005_Dcoef, Simonnin_2017_diffusion, Zaragoza_2019_TIP4P}. 

\begin{figure}[htbp]
    \centering
    \includegraphics[width=1.0\linewidth]{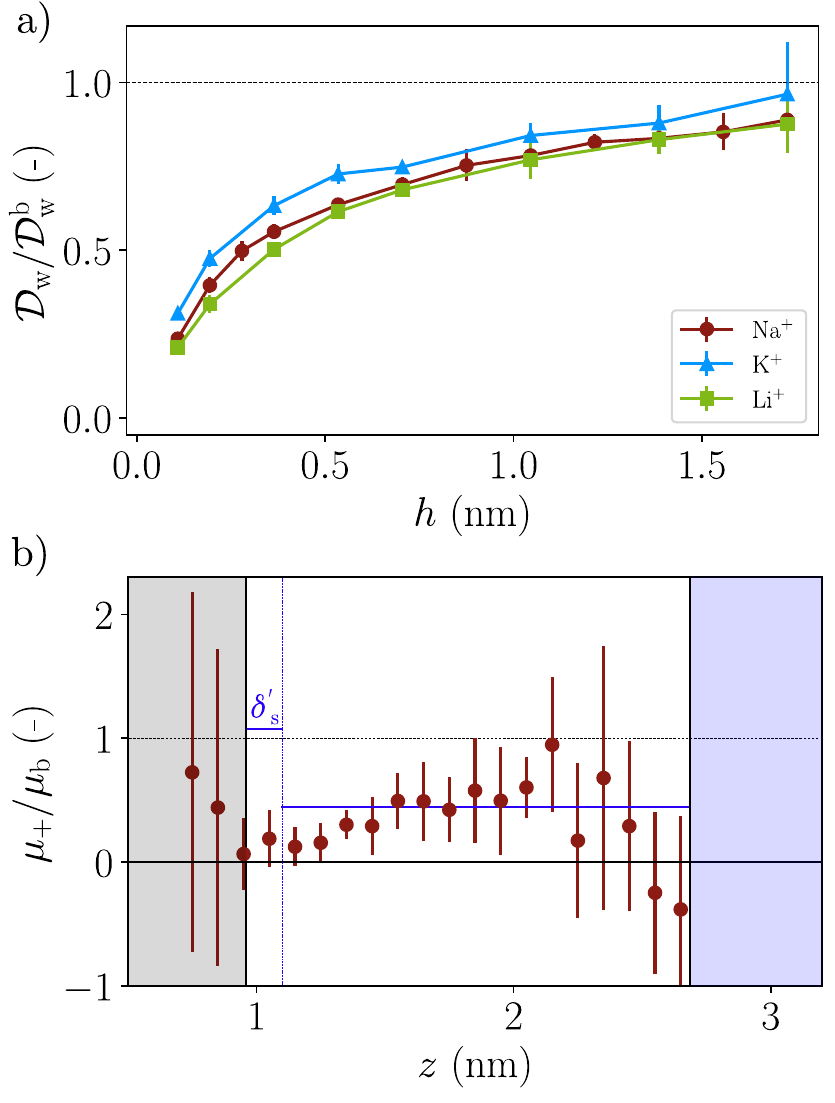}
    \caption{(a) Water average self-diffusion coefficient, normalized by its value in bulk \cite{Tazi_2012}, as a function of the wetting film thickness, for the three ion types. (b) Typical cation mobility profile $\mu_+(z)$ for Na$^+$ computed from equation \eqref{eq:mu_cat}. The solid line correspond to a theoretical rescaled constant mobility with the effective enhanced viscosity established in section \ref{sec:eo_vel}. 
    Both profiles profiles are extracted from a water film of thickness $h = \qty{1.7}{\nano\meter}$.} 
    \label{fig:water_D_mu_z}
\end{figure}

We first examine the diffusion properties of water molecules by computing their average self-diffusion coefficient, $\mathcal{D}_\text{w}$, parallel to the interface from their mean square displacement. 
We measured the diffusion coefficient from the mean squared displacement in the directions parallel to the interface. We did not apply finite size corrections: while finite size effects due to hydrodynamic interactions with periodic images can be large in simulations of bulk liquids, they are strongly reduced in the presence of a solid wall \cite{Simonnin_2017_diffusion,Zaragoza_2019_TIP4P}. 
Figure~\ref{fig:water_D_mu_z}(a) presents the variation of $\mathcal{D}_\text{w}$ with film thickness, normalized by its bulk value. 
{In our simulations, the computed bulk self-diffusion coefficient of water molecules is $\mathcal{D}_\text{w}^\text{b} = \left( 2.21 \pm 0.08 \right) \times 10^{-9} \, \si{\meter\squared\per\second}$ (see the \textit{supplementary material} for further details), that has to be compared to a value from the literature for pure water \cite{Tazi_2012} $\mathcal{D}_\text{w}^\text{b} = 2.49 \times 10^{-9}  \, \si{\meter\squared\per\second}$. The discrepancy between both values can be attributed to the presence of salts in our system at a concentration of $0.54 \, \si{\mol\per\liter}$. 
As the thickness of the wetting film increases, the effective cation concentration within the film decreases. For this reason, we normalize the water diffusion coefficient $\mathcal{D}_\mathrm{w}$ by the bulk diffusion coefficient of pure TIP4P/2005 water at $300 \, \si{\kelvin}$ reported in Ref.~\citenum{Tazi_2012}.}
{Water diffusion coefficient is essentially independent on the solute cation.}
{For the thickest films, $\mathcal{D}_\text{w}$ becomes close to the bulk value, {without exceeding it}.}
{The diffusion coefficient then }progressively decreases as the film thickness reduces, with a significant drop occurring in films thinner than one molecular layer.
Considering hydrodynamic boundary effects on diffusion \cite{Simonnin_2017_diffusion, Zaragoza_2019_TIP4P}, this behavior aligns with an intrinsic viscosity of the solvent that remains constant regardless of confinement. Specifically, it is expected that diffusion coefficients increase relative to bulk values at frictionless interfaces (e.g., liquid-vapor interfaces) and decrease near no-slip boundaries (e.g., liquid-solid interfaces). This interpretation is consistent with the simulations of Poiseuille flows, which indicated only a minor increase in effective viscosity, attributed primarily to ion-wall friction.

We then turn to the ion mobility profiles $\mu_+ (z)$. One uses the expression of the cation velocity under an electric field, given by:
\begin{equation}
    v_+(z) = v_x(z) + e \mu_+ (z) E_x
    \label{eq:mu_cat}
\end{equation}
with $v_+(z)$ and $v_x(z)$ respectively the cation and fluid velocity profiles (the latter one corresponds simply to the electro-osmotic velocity profile).
Figure~\ref{fig:water_D_mu_z}(b) shows the cation mobility profile for Na$^+$ and a water film thickness of $h=\qty{1.7}{\nano\meter}$. Typical mobility profiles for K$^+$ and Li$^+$ are shown on Figs.~S12 b) and c) in the \textit{supplementary material}. 

\begin{figure*}[htbp]
    \centering
    \includegraphics[width=1.0\linewidth]{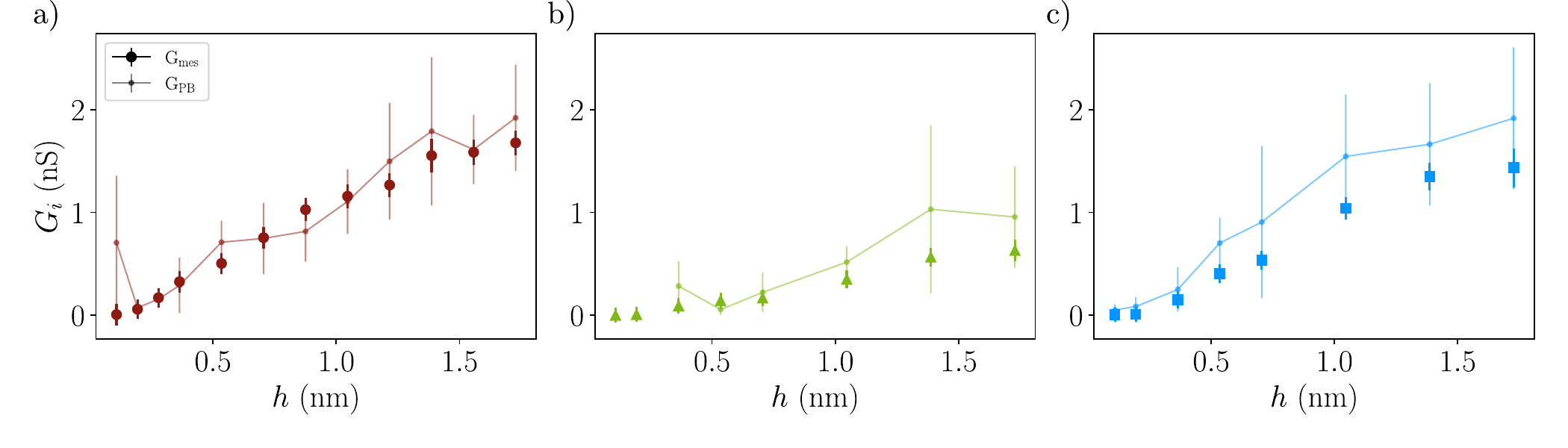}
    \caption{Numerical conductance $G$ as a function of the water film thickness $h$ for (a) Na$^+$, (b) K$^+$ and (c) Li$^+$. The filled marker corresponds to the numerical conductance and the open markers are the estimation of the conductance considering the concentration, electro-osmotic velocity and mobility profiles as discussed in the text.}
    \label{fig:G_est}
\end{figure*}

In contrast with water for which the effect is small, the mobility of ions is strongly reduced as compared to the bulk. The mobility vanishes close to the wall, consistently with the idea of a hindered diffusion layer. Additionally, in the rest of the film farther from the wall, the mobility is approximately constant, but is around twice lower than its bulk value for Na$^+$ and Li$^+$, and $\sim 4$ times lower for K$^+$. This decrease in ion mobility is well captured by a rescaled mobility using the effective enhanced viscosity found in electro-osmotic flows 
(see the colored solid lines in figure \ref{fig:water_D_mu_z} and Fig.~S12 from the \textit{supplementary material}). This is again consistent with the idea of an increased friction between the ions and the wall, especially for K$^+$, while the intrinsic viscosity of the solvent does not vary significantly. 
We note that the distance-dependent diffusion (or mobility) that we observe is expected  close to a solid boundary even for a homogeneous solvent viscosity, as a consequence of hydrodynamic hindrance by the wall \cite{Faxen1922,HappelBrenner1973,Saugey_2005_Dcoef,Siboulet2017}.

\subsection{Conductance estimation}

Thanks to the MD simulations, we have been able to have a look at the molecular details, especially at the cation distribution $c_+(z)$, fluid velocity $v_x(z)$ and cation mobility $\mu_+(z)$ profiles. These quantities are very important to estimate the overall conductance $G$, thanks to Eq.~\eqref{eq:G_contrib}. Figure~\ref{fig:G_est} shows the numerical conductance $G_\mathrm{mes}$ and the corresponding conductance estimation $G_\mathrm{PB}$ using the concentration, velocity and mobility profiles that have been determined previously.

One can see that for all cations, the estimated conductance give the same trend and the same amplitude of the numerical conductance, considering the errorbars. Overall, continuum descriptions adequately modified with information obtained from MD simulations can reproduce the global conductance of the system without empirical fitting parameters.

\section{Conclusion}

In this work, we have used molecular dynamics simulations to investigate ionic transport in ultra-confined wetting films on silica surfaces, for thicknesses ranging from $0.1$ to $\qty{1.7}{\nano\meter}$. Our simulations show that the electric conductance of the film increases with film thickness, in good agreement with recent experimental measurements~\cite{allemand_2023_cond}, and that the conductance depends strongly on the ion type. We then used the simulations to access molecular-level insights that are difficult to obtain experimentally. 

We first explored the distribution of the ions in the wetting films. At the solid-liquid interface, we observed a maximum of ion concentration shifted from the Gibbs dividing surface, explained by the the molecular roughness of the surface. At the liquid-vapor interface, ions are depleted due to image charge repulsion. Once these interfacial molecular layers are properly accounted for, the distribution of ions in the film is well described by the Poisson-Boltzmann theory. In that framework, one could compute an effective surface charge $\Sigma_\mathrm{PB}$, which was lower than the nominal surface charge, because part of the counter-ions lied in the molecular roughness region. Interestingly, $\Sigma_\mathrm{PB}$ was independent of the ion type, in contrast with the very different conductances observed, pointing to the critical role of dynamics. 

From a dynamical perspective, we show that electro-osmotic flows in these nanoconfined films are significantly reduced compared to continuum predictions using the nominal water viscosity. This reduction is captured by introducing an effective electro-osmotic viscosity, which is ion-specific, much larger than the bulk value, and reflects increased friction due to ion adsorption at the surface. Importantly, our results indicate that the intrinsic viscosity of water remains close to its bulk value, highlighting the specific role of interfacial effects and ion-surface coupling in the electrokinetic response.

The study further reveals pronounced ion-specific effects: K$^+$ ions exhibit notably lower conductance, stronger interfacial adsorption, and a larger effective viscosity than Na$^+$ and Li$^+$. This underlines the crucial role of ion nature and interfacial structure in governing transport at the nanoscale. 
With that regard, in experimental wetting films on silica, H$^+$ act as counter-ions, with interesting specific transport mechanism involving reactivity, such as the Grotthuss mechanism. These effects are not captured by standard non-reactive classical MD and are therefore left as an interesting perspective for future work. 

Finally, our results demonstrate that simple continuum-based one-dimensional models, suitably corrected to incorporate molecular-level effects identified through molecular dynamics simulations, remain effective tools for describing ionic conductance in nanometric and sub-nanometric wetting films. At film thicknesses comparable to the molecular size, which also align with typical surface roughness scales, a detailed molecular-level description naturally becomes more appropriate. Indeed, under these conditions, defining continuous fields perpendicular to the substrate becomes challenging, and lateral heterogeneities parallel to the interface must be explicitly considered. The \textit{supplementary material} provides data illustrating these heterogeneities, revealing the emergence of partially dewetted zones in extremely thin films. Importantly, these molecular-scale phenomena occur exclusively in ultra-thin films, for which the conductance inherently drops to zero, thereby preserving the validity and relevance of continuum approaches for practical purposes.

\section*{SUPPLEMENTARY MATERIAL}

The supplementary material includes additional information on the MD simulations, details on the theoretical description of the system (i.e. concentration profiles, EO velocity profiles, pressure induced flow profiles, ions trapping) and additional discussions on the mobility profiles and two dimensional effects. 

\section*{Acknowledgments}

The authors thank Benjamin Rotenberg for fruitful exchanges. This project was funded by the ANR project Soft Nanoflu (ANR-20-CE09-0025-03) and the IDEXLYON project from Lyon University in the program PIA (ANR-16-IDEX- 0005).

\section*{AUTHOR DECLARATIONS}

\textit{Conflict of interest}--- The authors have no conflict to disclose.

\section*{DATA AVAILABILITY}

Some data that support the findings of this study are openly available in Zenodo, at \url{https://doi.org/10.5281/zenodo.17724166}, reference number 17724166. Other data that support the findings of this study are available from the corresponding author upon reasonable request.

\providecommand{\noopsort}[1]{}\providecommand{\singleletter}[1]{#1}



\section*{Supplementary material}

# Supplementary materials: Molecular insight on ultra-confined ionic transport in wetting films: the key role of friction

Aymeric Allemand, Anne-Laure Biance, Christophe Ybert, and Laurent Joly*

*Université Claude Bernard Lyon 1, CNRS, Institut Lumière Matière, UMR5306, F69100 Villeurbanne, France*

# I. NUMERICAL DETAILS

## A. Wetting film system

One wetting film is composed of $N_w$ water molecules, with $N_w$ ranging from $N_w \simeq 10^3$ water molecules for a thickness $h \simeq 1.7\,\text{nm}$ to $N_w \simeq 60$ water molecules for a thickness $h \simeq 0.11\,\text{nm}$. This film is defined thanks to the software `Moltemplate` [1], which can create a large variety of systems. To avoid particle overlap, a cubic network is defined and ten cations are randomly added afterward. This system is then duplicated and combined with the silica wall.

## B. Silica substrate

Silica interfaces have been widely studied numerically [2, 3]. Here, crystalline and amorphous silica substrates have been considered, having similar dimensions of $4.200 \times 4.195 \times 2.330\,\text{nm}^3$ and $4.273 \times 4.480 \times 2.593\,\text{nm}^3$ respectively, with the same surface density of silanol groups $4.5\,\text{SiOH/nm}^2$. The crystalline substrate is smooth. The atomistic surface roughness of the amorphous substrate is determined thanks to the root-mean-squared deviation of the surface atoms from a plane determined by fitting to their positions [4], and is approximately $1.25\,\text{Å}$. The amorphous substrate has been extracted from the database of reference [4], whereas the crystalline solid as been generated by the web-based platform `CHARMM-GUI` that allows to interactively build complex systems [5, 6].

## C. Minimization step

An energy minimization step has been executed. The atomic coordinates are iteratively adjusted to minimize the overall energy of the system, following the minimization parameters: relative tolerance $10^{-4}$, force tolerance of $10^{-6}\,\text{kcal}\,\text{mol}^{-1}\,\text{Å}^{-1}$, maximum number of iterations set to $5.0 \times 10^4$ and a maximum number of energy evaluation of $5.0 \times 10^5$.

---

$^*$ laurent.joly@univ-lyon1.fr

| Atoms | $O_w$ | $H_w$ | M site | Na | K | Li |
|---|---|---|---|---|---|---|
| $\varepsilon$ (kJ mol$^{-1}$) | 0.774908 | 0 | 0 | 1.472356 | 1.985740 | 0.435090 |
| $\sigma$ (Å) | 3.15890 | 0 | 0 | 2.21737 | 2.30140 | 1.43970 |
| $q$ ($\times e$) | 0 | +0.5564 | −1.1128 | −0.85 | −0.85 | −0.85 |

TABLE S1. LJ parameters for the TIP4P/2005 water model and for cations.

### D. Force field parameters

We use the TIP4P/2005 water model [7]. This model is a rigid 4 sites model. The parameterization is based on simulating the water bulk properties and is summarized in table S1 for the Lennard-Jones (LJ) parameters and in table S2 for the harmonic bond and angle potentials. The *shake* algorithm [8] is used to rigidify bonds and angle. This model is one of the most robust and accurate [9] and has been widely used to study confined systems such as water in carbon nanotubes [10, 11].

To describe the charge carriers, we consider sodium (Na$^+$), potassium (K$^+$) and lithium (Li$^+$) cations, described with the Madrid model [12, 13]. This model implements in particular a scaled charge of 0.85 in electron units for the ions. This helps one to better simulate solution properties when combined with TIP4P/2005 water model such as densities up to high concentration, viscosity, or solubility. LJ parameters are summarized in table S1. Interactions with TIP4P/2005 water have been characterized by Zeron *et al.* and the parameters can be found in the corresponding article [12].

| Angle: $E_\theta = k_\theta \left(\theta_{ij} - \theta_{0,ij}\right)^2$ | | Bond: $E_r = k_r \left(r_{ij} - r_{0,ij}\right)^2$ | |
|---|---|---|---|
| Type | $k_\theta$ (kcal mol$^{-1}$) $\theta_{0,ij}$ (°) | Type | $k_r$ (kcal mol$^{-1}$ Å$^{-2}$) $r_{0,ij}$ (Å) |
| H-O-H | 75      104.52 | O-H | 600.0      0.9572 |

TABLE S2. Harmonic potentials for bond and angle for the TIP4P/2005 water model.

Silica interfaces are modeled with the Interface Force Field (IFF) [14], based on Lennard-Jones potential with parameters that are summarized in table S3. Equivalently to water molecules, harmonic bond, and harmonic angle potential are considered and table S4 summarize these potential parameters. The IFF has been shown to have good physical and chemical agreements with experimental data [15].

| Atoms | Si (deprotonated) | $O_b$ | $O_s$ (deprotonated) | H |
|---|---|---|---|---|
| $\varepsilon$ $(\mathrm{kJ\,mol^{-1}})$ | 0.38911 | 0.22594 | 0.51044 | 0.00000 |
| $\sigma$ (Å) | 3.70 | 3.09 | 3.09 | 0.00 |
| $q$ $(\times e$ C$)$ | 1.1 (0.819) | $-0.55$ | $-0.675$ $(-0.844)$ | 0.4 |

TABLE S3. LJ parameters for IFF $SiO_2$. The value within parentheses corresponds to the charge when the silanol group is deprotonated.

| Angle: $E_\theta = k_\theta\,(\theta_{ij} - \theta_{0,ij})^2$ | | | Bond: $E_r = k_r\,(r_{ij} - r_{0,ij})^2$ | | |
|---|---|---|---|---|---|
| Type | $k_\theta$ $(\mathrm{kcal\,mol^{-1}})$ | $\theta_{0,ij}$ (°) | Type | $k_r$ $(\mathrm{kcal\,mol^{-1}\,Å^{-2}})$ | $r_{0,ij}$ (Å) |
| O-Si-O | 100 | 109.5 | Si-O | 285 | 1.68 |
| Si-O-Si | 100 | 149.0 | O-H | 495 | 0.945 |
| Si-O-H | 50 | 115.0 | | | |

TABLE S4. Harmonic potentials for bond and angle for the silica surface.

To ensure overall system neutrality, it is necessary to adjust the charge of the silanol groups to be opposite to that of the cations. This adjustment can be achieved by solving the following set of equations:

$$\begin{cases} \Delta Q' = q'_{\mathrm{SiO^-}} - q_{\mathrm{SiOH}} = q'_{\mathrm{Si,d}} + q'_{\mathrm{O_s,d}} - q_{\mathrm{Si}} - q_{\mathrm{O_s}} - q_{\mathrm{H_s}} = -0.85 \\ \Delta Q' = k\,(q_{\mathrm{Si,d}} - q_{\mathrm{Si}} + q_{\mathrm{O_s,d}} - q_{\mathrm{O_s}}) - q_{\mathrm{H_s}} = -0.85 \end{cases} \quad (S4)$$

In these equations, $\Delta Q'$ represents the new total charge difference of the silanol groups, $q_{\mathrm{X}}$ denotes the individual charge of atom X (X = Si, $O_s$, $H_s$) [15], and $q'_{\mathrm{Y,d}}$ represents the new charge of the deprotonated species Y (Y = Si, $O_s$). The subscript "d" designates the species belonging to a deprotonated silanol group. The first equality of equation (S4) corresponds to the charge difference between the scaled charge of an ionized silanol group $q'_{\mathrm{SiO^-}}$ and the unchanged charge of a SiOH group. To find the scaled charge $q'_{\mathrm{SiO^-}}$, one has to assume that the difference between the charge of a silicon atom (respectively oxygen atom) of a $SiO^-$ group and a silicon atom (resp oxygen atom) of a SiOH group is decreased by a factor $k$: $q'_{\mathrm{Y,d}} - q_{\mathrm{Y}} = k\,(q_{\mathrm{Y,d}} - q_{\mathrm{Y}})$ with Y = Si, $O_s$. The second equality of equation (S4) takes into account this assumption. Finally, the values of $q'_{\mathrm{Si,d}}$ and $q'_{\mathrm{O_s,d}}$ can be approximated as 0.819 and $-0.844$ respectively.

### E. Simulation protocol

In the simulations that are carried out in this work, one uses the $(N, V, T)$ ensemble. Therefore, we need to control the temperature in these MD simulations: we use a Nosé-Hoover thermostat with a damping factor of $100\,\delta t$ [16].

Figure S1 shows a schematic of a typical simulation. Since the focus of the study is on the dynamics of the water film and the silica interface, the equation of motion is computed for the water molecules, cations, and silanol groups only.

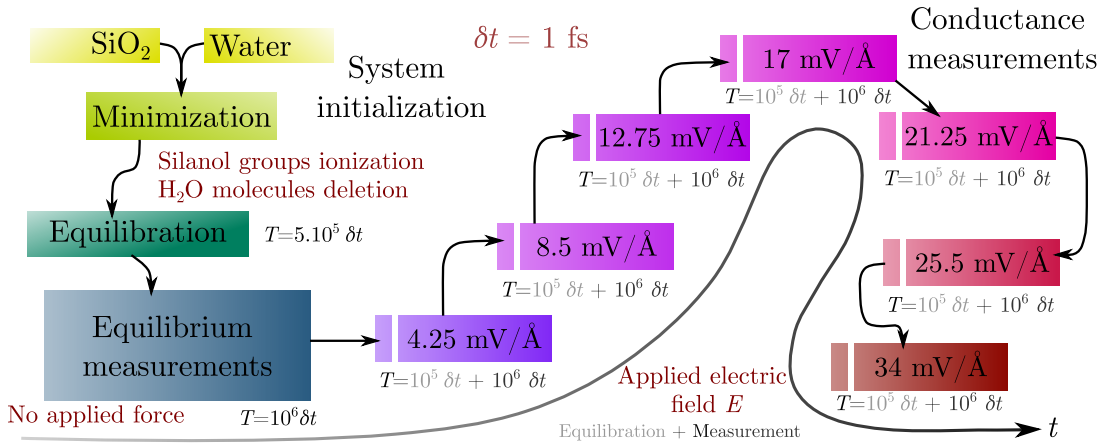

FIG. S1. Schematic of the steps of a simulation.

A single simulation consists of executing this entire simulation procedure, which has a total duration of 9.2 ns. However, in order to obtain more accurate estimates of the measured quantity and its error, the simulation is repeated three times. Specifically, for each interface, 10 silanol groups are randomly deprotonated in one simulation, and three random silanol group ionizations are performed.

### F. Thickness measurements

In this section, we provide additional details about the measurements of the water film thickness. In the simulations, the number of water molecules is constant and sets the thickness of this wetting film. To compute its thickness, a *Gibbs dividing plane* approach is used. This technique is widely used and discussed [17–19], to compute position $z_{sl}$ separating a region of homogeneous fluid and a region without fluid. In this case, this method is summarized by the following equation:

$$\int_0^{+\infty} \rho(z)\mathrm{d}z = \int_{z_{\mathrm{sl}}}^{z_{\mathrm{sl}}+h} \rho_{\mathrm{b}}\mathrm{d}z \tag{S5}$$

with $\rho(z)$ the mass density profile of the wetting film and $\rho_{\mathrm{b}}$ the mass density in the bulk. We neglect, here, the vapor mass density and consider that all water molecules are located in the wetting films. In fact, during one simulation, only a few water molecules are observed in the vapor phase. Figure S2 shows the mass density profile as a function of the $z$ coordinate. Equation (S5) corresponds mathematically to the equality of the purple and the green areas.

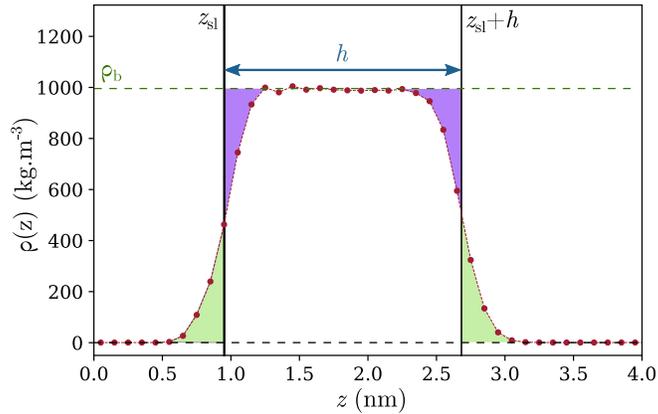

FIG. S2. Example of a mass density profile of a wetting film as a function of vertical position $z$. This figure illustrates the thickness measurement using Gibbs dividing plane approach summarized by equations (S5) or (S6). Mathematically, the first equation is the equality between the green and the purple areas. Here, the thickness is equal to $h = 1.73\,\mathrm{nm}$.

One can note that $\int_{z_{\mathrm{sl}}}^{z_{\mathrm{sl}}+h} \rho_{\mathrm{b}}\mathrm{d}z = \rho_{\mathrm{b}}h$ and that $\int_{-\infty}^{+\infty} \rho(z)\mathrm{d}z = N_{\mathrm{w}}m_{\mathrm{H_2O}}/L_x L_y$ with $L_i$ the size of the simulation box in the direction $i = x$ or $y$, $N_{\mathrm{w}}$ the number of water molecules in the wetting film and $m_{\mathrm{H_2O}}$ the mass of a single water molecule. This leads to the following simple equation to measure the water film thickness:

$$\rho_{\mathrm{b}}h L_x L_y = N_{\mathrm{w}}m_{\mathrm{H_2O}}. \tag{S6}$$

Numerically, we used equation (S6) to compute the thickness of the wetting film $h$, and thanks to this value, we could use equation (S5) to fit a rectangular function to the water density profile, to estimate the position $z_{\mathrm{sl}}$: $|z_{\mathrm{sl}}| = 0.96 \pm 0.01\,\mathrm{nm}$ for $\mathrm{Na^+}$, $|z_{\mathrm{sl}}| = 0.98 \pm 0.01\,\mathrm{nm}$ for $\mathrm{K^+}$, and $|z_{\mathrm{sl}}| = 0.98 \pm 0.01\,\mathrm{nm}$ for $\mathrm{Li^+}$.

## II. THEORETICAL FRAMEWORK

In this section, we will describe the theoretical framework that has been used in this work to describe the profiles along the $z$-axis of the electric potential, the cation concentration, the electro-osmotic velocity and the ion mobility.

### A. Concentration profile

We consider the system described by the inset of Fig. 1 in the main text. The coordinate system is the one from the MD simulations with $z = 0$, the center of the solid medium. The solid/liquid interface is at the position $z_{\mathrm{sl}}$ and we define $z_0 = z_{\mathrm{sl}} + \delta$ the position at which the electric potential is set to its surface value $V_s$, i.e. where the cation concentration profile reaches its maximum value. The vapor-liquid interface position is located at $z = z_{\mathrm{sl}} + h$.

As described in the main text, one can see that cations are excluded from the liquid-vapor interface. This phenomenon has been originally described in an electrolyte [20]. In our case, we do not have added salt: cations are released from the surface, the so-called Gouy-Chapman regime. To describe this exclusion layer, we introduce a boundary layer of thickness $\delta_e$ beyond which the cation profile is zero at $z = z_{\mathrm{sl}} + h - \delta_e$. We use a concentration threshold criteria $\delta c_{\mathrm{th}}$ to set the value of $\delta_e$. Figure S3 shows a typical concentration profile with the boundary of this exclusion layer for six difference threshold values. We fixed this value to $\delta c_{\mathrm{th}} = 0.15 \, \mathrm{mol/L}$, because based on the fitting curves, it is the best one to represent the exclusion layer.

Between $z_{\mathrm{sl}} + \delta$ and $z_{\mathrm{sl}} + h - \delta_e$, we use the PB equation to model the concentration profile:

$$\Delta V = -\frac{e c_{\mathrm{ref}}}{\varepsilon} \exp\left(-\frac{eV}{k_{\mathrm{B}}T}\right) \quad \Leftrightarrow \quad \frac{e}{k_{\mathrm{B}}T} V''(z) = -2K^2 \exp\left(-\frac{eV(z)}{k_{\mathrm{B}}T}\right), \qquad (\mathrm{S7})$$

with $V$, the electrical potential, $e$ the elementary charge, $\varepsilon$ the dielectric permittivity, $T = 300$ K the temperature, $k_{\mathrm{B}}$ the Boltzmann constant, $c_{\mathrm{ref}}$ a reference concentration for which the potential is zero and $K^{-1}$ a scale length defined as $K^2 = 2\pi \ell_B c_{\mathrm{ref}}$.

We consider that at $z = z_{\mathrm{sl}} + \delta$, we have a constant certain surface charge density $\Sigma_{\mathrm{PB}}$, and we set the electric potential to zero at $z = z_{\mathrm{sl}} + h - \delta_{\mathrm{e}}$. These boundary conditions are given by the following system:

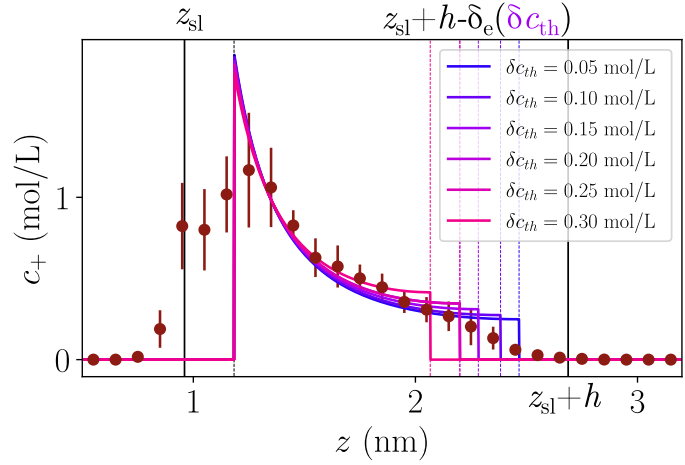

FIG. S3. Typical cation concentration profile for $Na^+$ ($h = 1.73\,\text{nm}$) as a function of $z$-coordinate. Fitting curves are shown as solid lines for six different values of concentration threshold $\delta c_{\text{th}}$.

$$
V'(z = z_{\text{sl}} + \delta) = \frac{4\pi\ell_{\text{B}}\Sigma_{\text{PB}}}{e},
$$
$$
V(z = z_{\text{sl}} + h - \delta_{\text{e}}) = 0,
$$
$$
V(z) = 0 \quad \forall z \notin [z_{\text{sl}} + \delta, z_{\text{sl}} + h - \delta_{\text{e}}].
$$
(S7)

The solution of equation (S7) in the limit of Gouy-Chapman is given by:

$$
V(z) = \frac{k_{\text{B}}T}{e}\log\left(\cos^2\left(K(z - h - \delta_{\text{e}})\right)\right),
$$
$$
K(h - \delta - \delta_{\text{e}})\tan\left(K(h - \delta - \delta_{\text{e}})\right) = \frac{h - \delta - \delta_{\text{e}}}{\ell_{\text{GC}}}.
$$
(S8)

The concentration profile is then given by a Boltzmann distribution:

$$
c_+(z) = c_{\text{ref}}\exp(-eV(z)/k_{\text{B}}T).
$$

Typical concentration profiles for six different water film thickness are shown on figures S4 for $Na^+$, S5 for $K^+$, and S6 for $Li^+$. The fitting curves are in good agreement with the numerical data for large water film thicknesses.

The variation of the parameters $\delta$ and $\delta_{\text{e}}$, and $\Sigma_{\text{PB}}$ as a function of $h$ are presented in Fig. 4(a), (b) and (c) of the main text.

## B. Electro-osmotic velocity profile

An electric field $E$ applied along the $x$ direction parallel to the interface induces a coulombic force on the non-neutral part of the fluid. A flow thus appears and is known as the

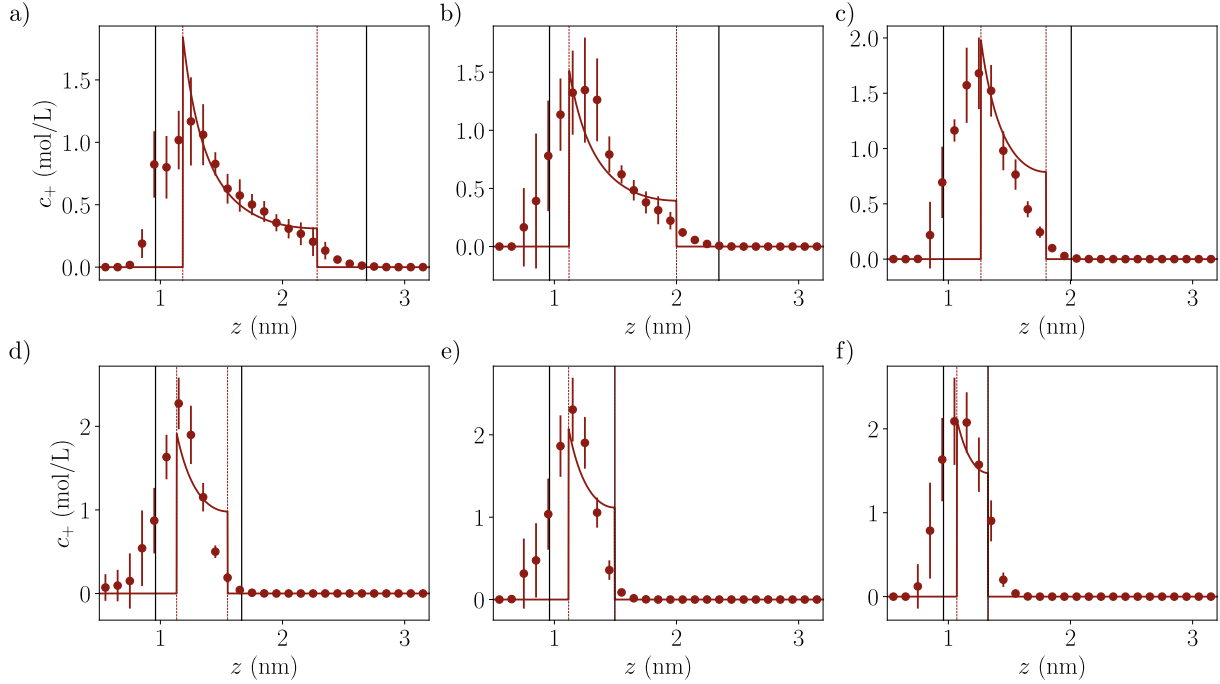

FIG. S4. Ion concentration profiles as a function of the $z$-position for Na$^+$ and six different water film thicknesses $h$: (a) $h = 1.73$ nm, (b) $h = 1.39$ nm, (c) $h = 1.05$ nm, (d) $h = 0.71$ nm, (e) $h = 0.54$ nm and (f) $h = 0.37$ nm. The solid vertical black lines delimit the wetting film according to Gibbs dividing plane method, i.e. $z = z_{\mathrm{sl}}$ and $z = z_{\mathrm{sl}} + h$ and the dashed red lines correspond respectively from left to right to $z = z_{\mathrm{sl}} + \delta$ and $z = z_{\mathrm{sl}} + h - \delta_{\mathrm{e}}$.

electro-osmotic flow [21]. The theoretical framework used to describe the electro-osmotic velocity profile $v$ along the $x$ direction is detailed in this section. According to the system symmetry, the velocity profile depends only on the $z$-coordinate (i.e. perpendicular direction to the interface). This velocity $v$ is given by solving Stokes equation, using Coulombic force:

$$\eta \frac{\mathrm{d}^2 v}{\mathrm{d}z^2} = -e c_+ E \tag{S9}$$

with $\eta$ the fluid viscosity. We use a stagnant layer of thickness $\delta_{\mathrm{s}}$ close to the solid-liquid interface, i.e., $v = 0$ for $z < z_{\mathrm{sl}} + \delta_{\mathrm{s}}$. One then needs to distinguish to cases: $\delta_{\mathrm{s}} \geq \delta$ and $\delta_{\mathrm{s}} \leq \delta$.

Let's solve the velocity profile for $\delta_{\mathrm{s}} \leq \delta$. One first needs to divide the film into four region:

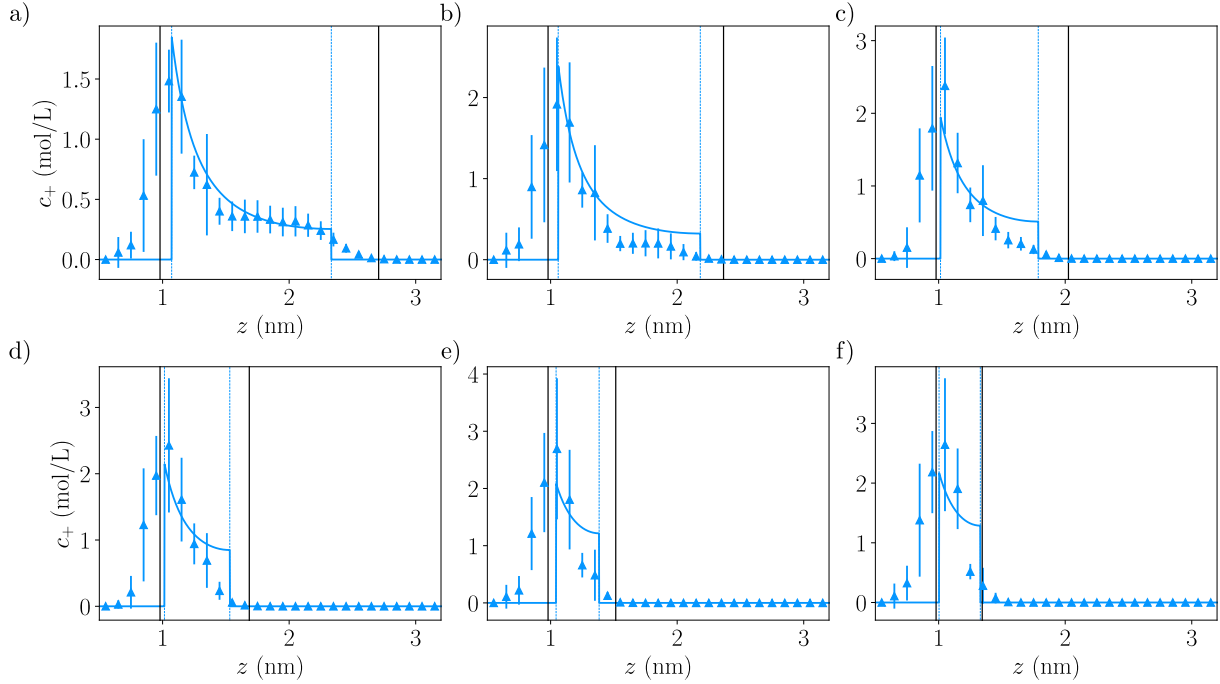

FIG. S5. Ion concentration profiles as a function of the $z$-position for K$^+$ and and six different water film thicknesses $h$: (a) $h = 1.73$ nm, (b) $h = 1.39$ nm, (c) $h = 1.05$ nm, (d) $h = 0.71$ nm, (e) $h = 0.54$ nm and (f) $h = 0.37$ nm. The solid vertical black lines delimit the wetting film according to Gibbs dividing plane method, i.e. $z = z_{\text{sl}}$ and $z = z_{\text{sl}} + h$ and the dashed blue lines correspond respectively from left to right to $z = z_{\text{sl}} + \delta$ and $z = z_{\text{sl}} + h - \delta_{\text{e}}$.

$$
\begin{aligned}
&\text{Region I}: z_{\text{sl}} \leq z \leq z_{\text{sl}} + \delta_{\text{s}} &&\Leftrightarrow && v(z) = 0 \\
&\text{Region II}: z_{\text{sl}} + \delta_{\text{s}} \leq z \leq z_{\text{sl}} + \delta &&\Leftrightarrow && \frac{\text{d}^2 v}{\text{d} z^2} = 0 \\
&\text{Region III}: z_{\text{sl}} + \delta \leq z \leq z_{\text{sl}} + h - \delta_{\text{e}} &&\Leftrightarrow && \eta \frac{\text{d}^2 v}{\text{d} z^2} = -e c_+(z) E \\
&\text{Region IV}: z_{\text{sl}} + h - \delta_{\text{e}} \leq z \leq z_{\text{sl}} + h &&\Leftrightarrow && \frac{\text{d}^2 v}{\text{d} z^2} = 0
\end{aligned}
$$

One can use Poisson equation given by $\varepsilon \Delta V(z) = -e c_+(z)$ and integrate this equation twice to solve the EO velocity profile. This gives:

$$
\begin{aligned}
&v(z) = 0 &&\text{for I} \\
&v(z) = \alpha \left( z - z_{\text{sl}} - \delta_{\text{s}} \right) &&\text{for II} \\
&v(z) = \frac{\varepsilon}{\eta} \left( V(z) - V(z_{\text{sl}} + \delta) \right) + \alpha \left( \delta - \delta_{\text{s}} \right) &&\text{for III} \\
&v(z) = \beta &&\text{for IV}
\end{aligned}
\tag{S10}
$$

with $\alpha = \frac{\varepsilon}{\eta} \partial_z V(z_{\text{sl}} + \delta) E_x$ and $\beta = \alpha(\delta - \delta_{\text{s}}) - \frac{\varepsilon}{\eta} V(z_{\text{sl}} + \delta)$, two constants that have

x

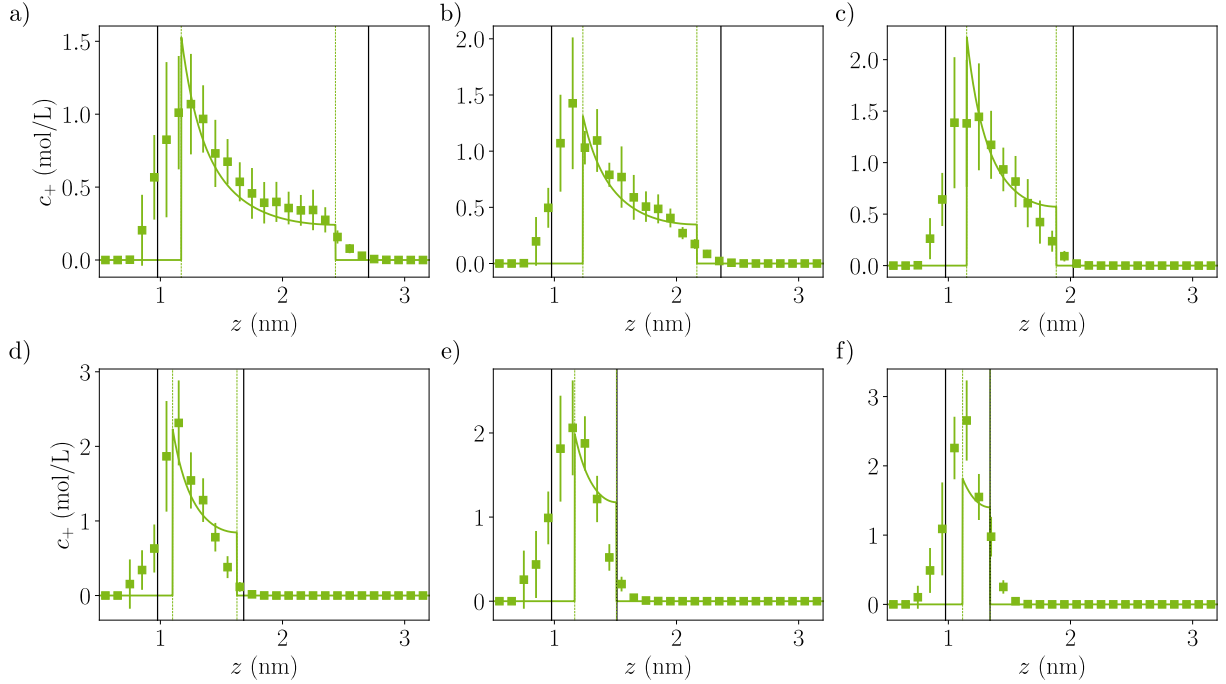

FIG. S6. Ion concentration profiles as a function of the $z$-position for Li$^+$ and and six different water film thicknesses $h$: (a) $h = 1.73$ nm, (b) $h = 1.39$ nm, (c) $h = 1.05$ nm, (d) $h = 0.71$ nm, (e) $h = 0.54$ nm and (f) $h = 0.37$ nm. The solid vertical black lines delimit the wetting film according to Gibbs dividing plane method, i.e. $z = z_{sl}$ and $z = z_{sl} + h$ and the dashed green lines correspond respectively from left to right to $z = z_{sl} + \delta$ and $z = z_{sl} + h - \delta_{e}$.

been determined with the continuity of the velocity and its derivative between regions, i.e. $v_i = v_{i+1}$ and $\partial_z v_i = \partial_z v_{i+1}$, with $i =$I, II or III.

Following an equivalent path, the velocity field for $\delta_s \geq \delta$ is given by:

$$
\begin{aligned}
v(z) &= 0 & \text{for } z_{sl} \leq z \leq z_{sl} + \delta_s \\
v(z) &= \frac{\varepsilon}{\eta} \left( V(z) - V(z_{sl} + h - \delta_{e}) \right) & \text{for } z_{sl} + \delta_s \leq z \leq z_{sl} + h - \delta_{e} \\
v(z) &= \frac{\varepsilon}{\eta} V(z_{sl} + \delta_s) & \text{for } z_{sl} + h - \delta_{e} \leq z \leq z_{sl} + h
\end{aligned}
\tag{S11}
$$

It is important to note that the permittivity does not appear in the Stokes equation (S9). This means that the velocity profiles from equations (S10) and (S11) depends on $\varepsilon$ only through the use of Poisson equation. Actually, the EO velocity field can be expressed as the double integral of the charge density profile $\rho_{e} = ec_+$ [22]. That is why the viscosity $\eta_{EO}$ and the stagnant layer thickness $\delta_s$ have been used as fitting parameters from equations

(S10) and (S11) for the numerical EO velocity profile $v(z)$. Figures S7, S8 and S9 show the numerical velocity and its fitting curve as a function of $z$-coordinate for six different water film thicknesses for respectively Na$^+$, K$^+$ and Li$^+$.

The variation of the parameters $\eta_{\text{EO}}$ and $\delta_{\text{s}}$ as a function of $h$ are presented in Fig. 6(a) and (b) of the main text.

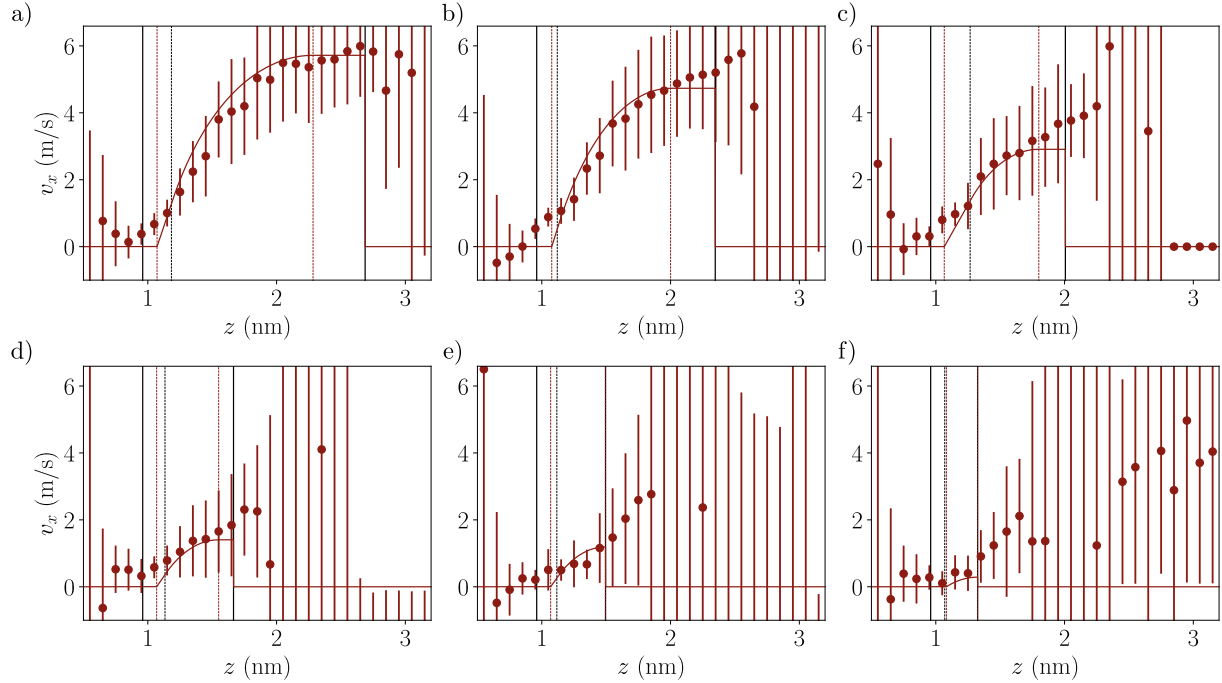

FIG. S7. Electro-osmotic velocity profiles as a function of the $z$-position for Na$^+$ and six different water film thicknesses $h$: (a) $h = 1.73$ nm, (b) $h = 1.39$ nm, (c) $h = 1.05$ nm, (d) $h = 0.71$ nm, (e) $h = 0.54$ nm and (f) $h = 0.37$ nm. The solid vertical black lines delimit the wetting film according to Gibbs dividing plane method, i.e. $z = z_{\text{sl}}$ and $z = z_{\text{sl}} + h$, the dashed red lines correspond respectively from left to right to $z = z_{\text{sl}} + \delta_{\text{s}}$ and $z = z_{\text{sl}} + h - \delta_{\text{e}}$ and the dashed black line is the limit $z = z_{\text{sl}} + \delta$.

## C.  Pressure-induced velocity profile

In order to investigate the increase in fluid viscosity, we simulated Hagen-Poiseuille flow in this wetting film system. With MD simulations, we can apply a force per atom $f_i$ to simulate pressure-driven flow. Using an equivalent system, we apply such force on the oxygen atoms

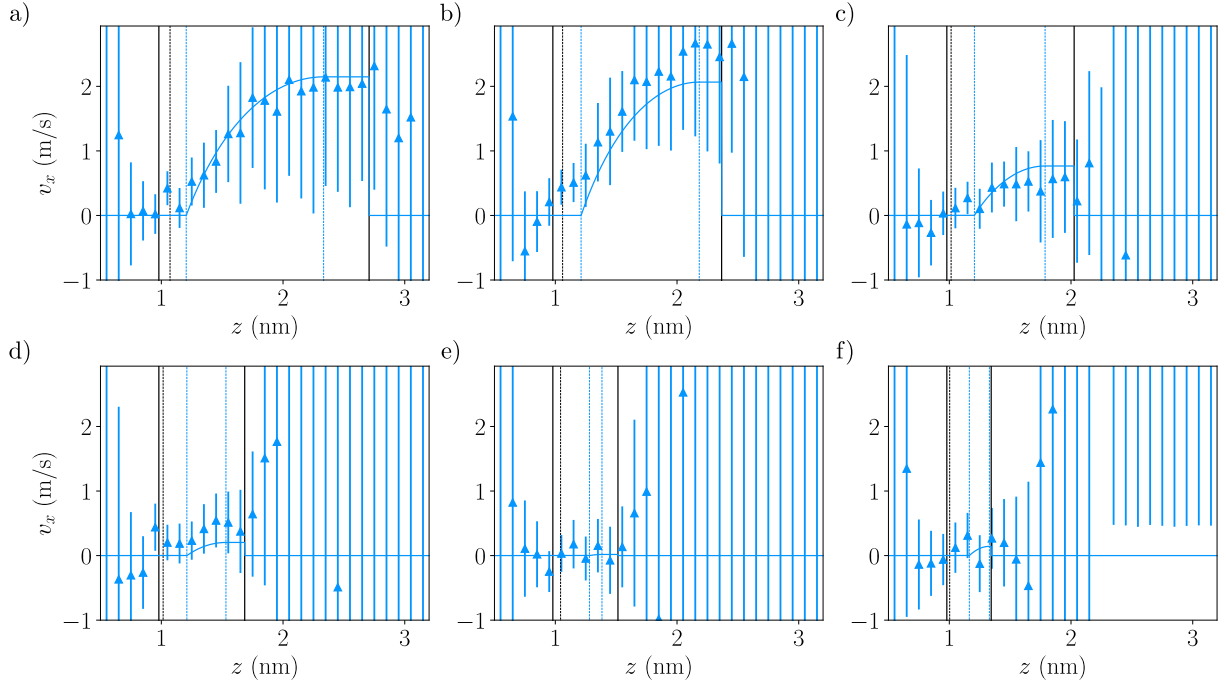

FIG. S8. Electro-osmotic velocity profiles as a function of the $z$-position for $K^+$ and six different water film thicknesses $h$: (a) $h = 1.73$ nm, (b) $h = 1.39$ nm, (c) $h = 1.05$ nm, (d) $h = 0.71$ nm, (e) $h = 0.54$ nm and (f) $h = 0.37$ nm. The solid vertical black lines delimit the wetting film according to Gibbs dividing plane method, i.e. $z = z_{sl}$ and $z = z_{sl} + h$, the dashed blue lines correspond respectively from left to right to $z = z_{sl} + \delta_s$ and $z = z_{sl} + h - \delta_e$ and the dashed black line is the limit $z = z_{sl} + \delta$.

of water molecules, ranging from 2.5 cal/(mol·Å) to 8.0 cal/(mol·Å).

Theoretically, this system can be solved using Stokes equation:

$$-\eta \frac{\mathrm{d}^2 v}{\mathrm{d}z^2} = f \tag{S13}$$

with $f$ a force density given by $f = n_b f_i = \Delta p / L_x$ with $n_b$, the fluid bulk density, $\Delta p$ the equivalent applied pressure difference and $L_x$ the system dimension along the $x$-direction. The range of applied force per atom corresponds to an applied pressure difference ranging from 24 MPa to 78 MPa. We checked that the results were in the linear response regime, with a mass flow rate proportional to the applied pressure difference, see figure S10. Figure S10 shows the mass flux $Q$ as a function of the equivalent applied pressure difference $\Delta p$. The results indicate that the system remains within the linear response regime over the explored range of $\Delta p$.

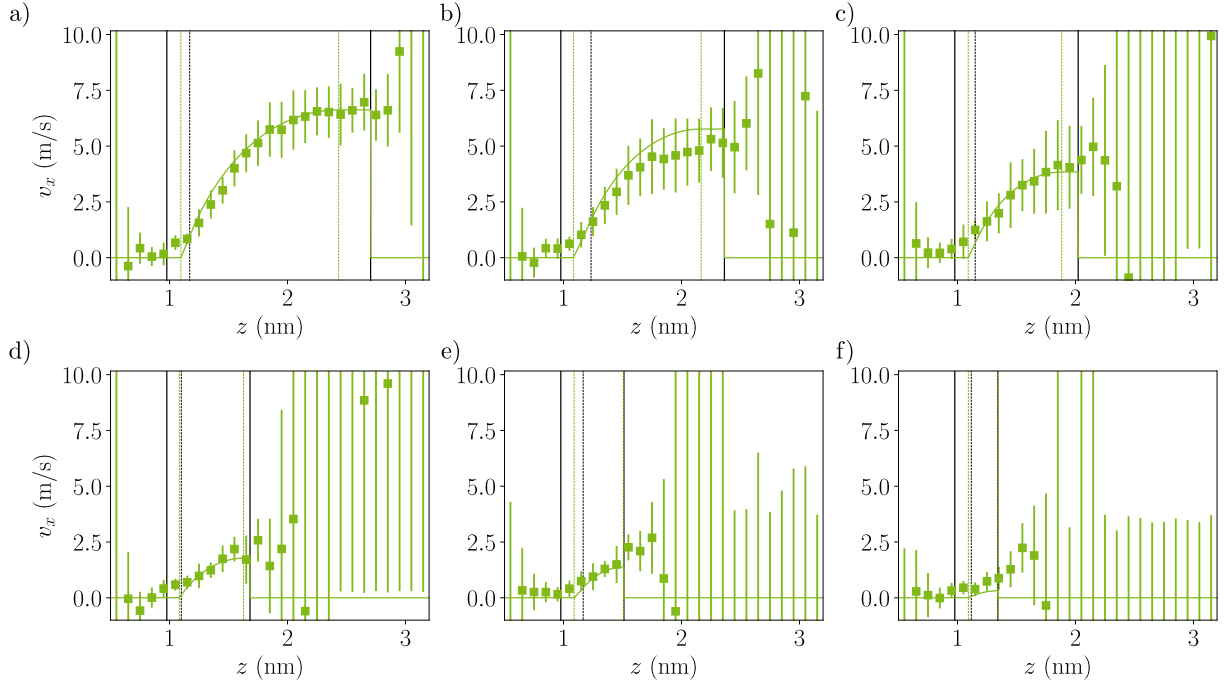

FIG. S9. Electro-osmotic velocity profiles as a function of the $z$-position for $Li^+$ and and six different water film thicknesses $h$: (a) $h = 1.73$ nm, (b) $h = 1.39$ nm, (c) $h = 1.05$ nm, (d) $h = 0.71$ nm, (e) $h = 0.54$ nm and (f) $h = 0.37$ nm. The solid vertical black lines delimit the wetting film according to Gibbs dividing plane method, i.e. $z = z_{sl}$ and $z = z_{sl} + h$, the dashed green lines correspond respectively from left to right to $z = z_{sl} + \delta_s$ and $z = z_{sl} + h - \delta_e$ and the dashed black line is the limit $z = z_{sl} + \delta$.

In the end, one gets:

$$v(z) = \frac{n_{\mathrm{b}} f_i}{2\eta} \left( \left( h - \delta_{\mathrm{s}}^{\mathrm{P}} \right)^2 - (z_{sl} + h - z)^2 \right) \tag{S13}$$

where we assumed $v\left(z \leq z_{sl} + \delta_{\mathrm{s}}^{\mathrm{P}}\right) = 0$, and no friction at the water/vapor interface, i.e. $\frac{\partial v}{\partial z}\big|_{z_{sl}+h} = 0$. This velocity profile is a Hagen-Poiseuille-like flow profile with a parabolic behavior, as expected. Fig S11 show the pressure-induced flow as a function of the $z$-coordinate for five values of applied force $f_i$. The solid lines correspond to fitting curves using equation (S13), and two fitting parameters: $\eta_{\mathrm{P}}$ and $\delta_{\mathrm{s}}^{\mathrm{P}}$ (see table II of the main article for the parameters value).

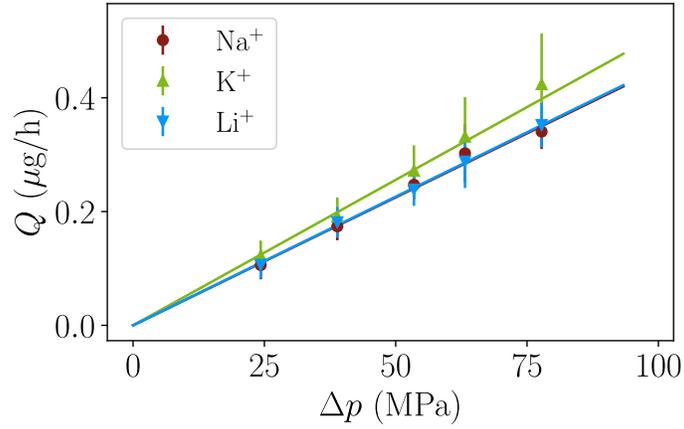

FIG. S10. Mass flow rate Q as a function of the applied pressure difference $\Delta p$ for $h = 1.73\,\mathrm{nm}$. The solid lines corresponds to linear fit.

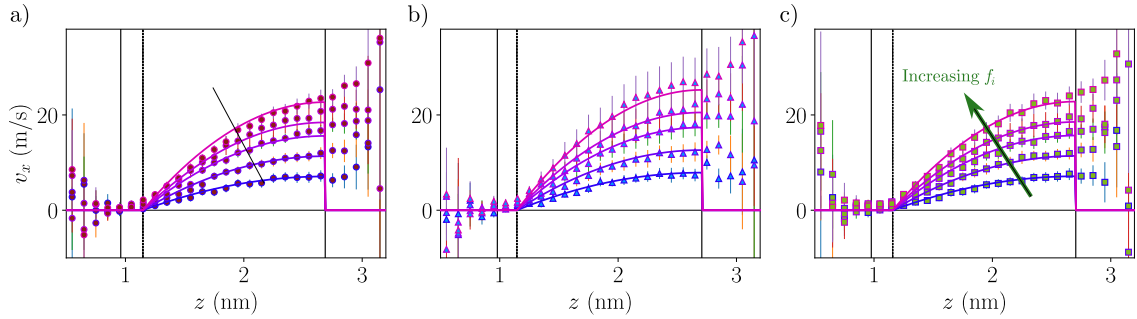

FIG. S11. Hagen-Poiseuille velocity profiles as a function of the $z$-position for (a) Na$^+$, (b) K$^+$ and (c) Li$^+$ The solid vertical black lines delimit the wetting film according to Gibbs dividing plane method, i.e. $z = z_{\mathrm{sl}}$ and $z = z_{\mathrm{sl}} + h$, the dashed green lines correspond respectively from left to right to $z = z_{\mathrm{sl}} + \delta_{\mathrm{s}}$ and $z = z_{\mathrm{sl}} + h - \delta_{\mathrm{e}}$ and the dashed black line is the limit $z = z_{\mathrm{sl}} + \delta$.

## D. Mobility profile

Ions in a liquid subjected to an electric field will move relative to the liquid, with a steady state relative velocity proportional to the applied electric force. The proportionality coefficient is the mobility, denoted $\mu$. Accordingly, in our MD simulations, the ion mobility profile $\mu_+(z)$ has been computed using the velocity of the cation $v_+$:

$$v_+(z) = v_{eo}(z) + e\mu_+(z)E_x \tag{S15}$$

with $v_{eo}$ the electro-osmotic velocity profile and $E_x$ the applied electric field. Figure S12 shows typical mobility profiles for (a) Na$^+$, (b) K$^+$ and (c) Li$^+$.

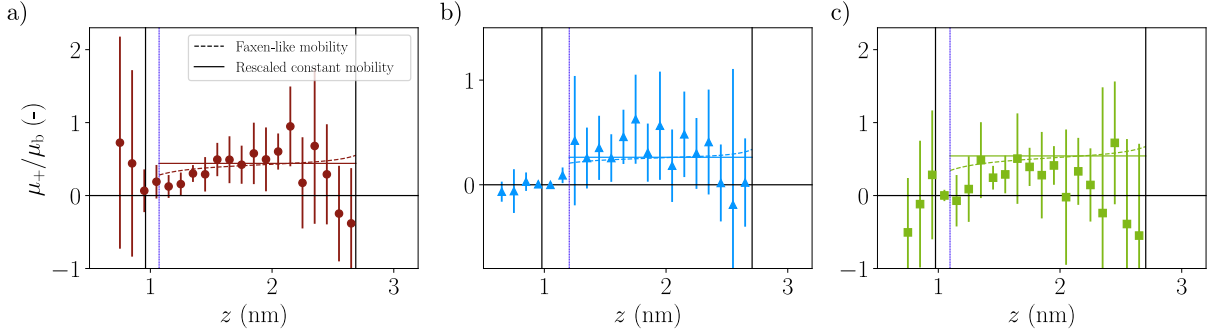

FIG. S12. Mobility profile for (a) Na$^+$ (b) K$^+$ and (c) Li$^+$ as a function of the $z$ coordinate.

One can see that the mobility profile is not equal to the bulk mobility $\mu_{\rm b}$. Moreover, the mobility profile is almost zero at both interfaces. This is due to cation exclusion at the free surface, i.e. liquid-vapor interface and due to strong friction and substrate roughness at the solid-liquid interface. The vanishing cations mobility close to the solid-liquid supports the additional friction force and the enhance overall viscosity.

Equation (S16) is the well-known Stokes-Einstein equation and allows us to compute an averaged cation mobility $\overline{\mu_+}$ as a function of the decreased viscosity:

$$\overline{\mu_+} = \frac{1}{3\pi\eta d_{\rm h}} \tag{S16}$$

with $d_{\rm h}$ the cation hydrodynamic diameter, assumed to be constant here.

Figure S12 show the good estimation of the averaged mobility $\overline{\mu_+}$ from equation (S16) as a solid line. It is worth mentioning that a Faxen-like mobility profile have been considered (shown in dashed line in figure S12) [9, 23] but no major improvement has been observed. Additionally, Faxen-like mobility is valid for $d_h \ll h$, that can be challenged in our case.

## III. ION TRAPPING

We computed the average number of adsorbed cations based on a static criterion, namely, we counted the number of cations that are located in the first neighbor shell of the charged groups. To quantify the number of cations located in the first neighbor shell, we first need to quantify the first neighbor distance.

## A. Radial density function

The radial distribution function (RDF) $g(r)$ describes how the local density of species $k$ varies with respect to the distance $r$ from a reference particle of species $j$. In a liquid system, the RDF plays a crucial role in identifying the local arrangement of molecules and the position of nearby neighbors, including the first, second, or third coordination shells. This function is particularly useful for characterizing the hydration shell of an ion and understanding its solvation in the liquid environment. Figure S13 presents the RDF between ionized silanol groups and three cations: Na$^+$, K$^+$, and Li$^+$, considering a water film thickness of $h = 1.7\,\text{nm}$.

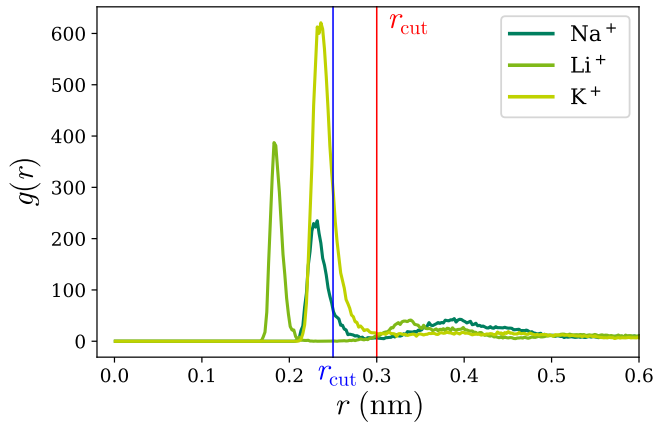

FIG. S13. Radial distribution function $g(r)$ as a function of the radial position $r$ for ionized silanol groups and considered cations (i.e. Na$^+$, Li$^+$, and Li$^+$). The considered water film has a thickness of $h = 1.7\,\text{nm}$ on an amorphous silica substrate.

For all three cations, the RDF $g(r)$ exhibits a similar overall shape, but there are slight differences in the amplitudes and positions of the first peak. Specifically, the position of the first peak, denoted as $r_1$, is the smallest for Li$^+$ at $1.83\,\text{Å}$, followed by Na$^+$ at $2.30\,\text{Å}$, and K$^+$ at $2.35\,\text{Å}$. These variations in position can be attributed to the differences in ion size, as Li$^+$ is the smallest cation while K$^+$ is the largest among those considered.

## B. Adsorbed ions

RDF measurements allow one to quantify the number of ions that are adsorbed, denoted as $N$, by setting a first neighbor threshold, manually chosen as the first non-zero minimum

of $g(r)$, noted $r_{cut}$ and shown in red for $Na^+$ and $K^+$, and in blue for $Li^+$ in figure S13. If the distance satisfies this criterion, the cation is considered as adsorbed at the interface. Figure 6 in the main article presents the proportion of trapped ions $N/N_{tot}$ as a function of the water film thickness $h$.

To ensure statistical significance, this measurement has been computed at regular intervals of every 2000 timesteps, and the results have been time-averaged as well as ensemble-averaged over the six simulated water films.

## IV.  FILM THICKNESS HETEROGENEITY

In the theoretical framework, the liquid is assumed to have a homogeneous thickness. Yet the substrate is chemically heterogeneous and especially for the thinnest films the homogeneity of the water molecules distribution is to be tested.

### A.  Water film structure

To begin with, one can examine the structural characteristics of water. Figure S14(a) displays a snapshot of a simulation with $N_w = 162$ water molecules and 10 sodium cations, representing a global water film thickness of $h = 0.27\,\text{nm}$. The snapshot reveals that the water molecules do not fully cover the silica interface, indicating the existence of dry spots. Additionally, the water molecules tend to cluster and overlap on the interface instead of being uniformly distributed. To quantify relevant properties like the mass density, the $(x, y)$ planes are divided into a grid with spacing $\delta x \simeq \delta y \simeq 1.1\,d_{H2O}$, where $d_{H2O}$ represents the diameter of a water molecule. Figure S14(b) shows the film local thickness $H(x, y)$ normalized by $h$ estimated from the ratio of the local water density and the bulk density $\rho(x, y)/\rho_b$. The water mass density is computed using the following expressions:

$$\begin{cases} \rho(x_i, y_i, t) = \sum_k \dfrac{m_{H2O}}{h\,\delta x \delta y} \delta\left(\mathbf{r}_i - \mathbf{r}_k\right) \\ \rho(t) = \sum_{x_i} \sum_{y_i} \rho(x_i, y_i, t) = \rho_b \end{cases} \tag{S18}$$

In the first expression, the mass density is obtained by summing the contributions of water molecules located within the grid element centered at $(x_i, y_i)$. The second equality of equation (S18) is a direct consequence of the definition of $h$. The grid element size is given by $\delta x$

and $\delta y$. The ratio between the time-averaged density and the bulk density $\rho(x_i, y_i)/\rho_{\rm b}$ gives an estimation of the local water film thickness $H(x_i, y_i)$ compared to the measured one $h$. To obtain an averaged value, the density within each grid element is computed every 2000 timesteps and then averaged over time. This procedure is consistent with the measurement of mass density. Additionally, by using this grid, one can also determine the number of water molecules present within each grid element.

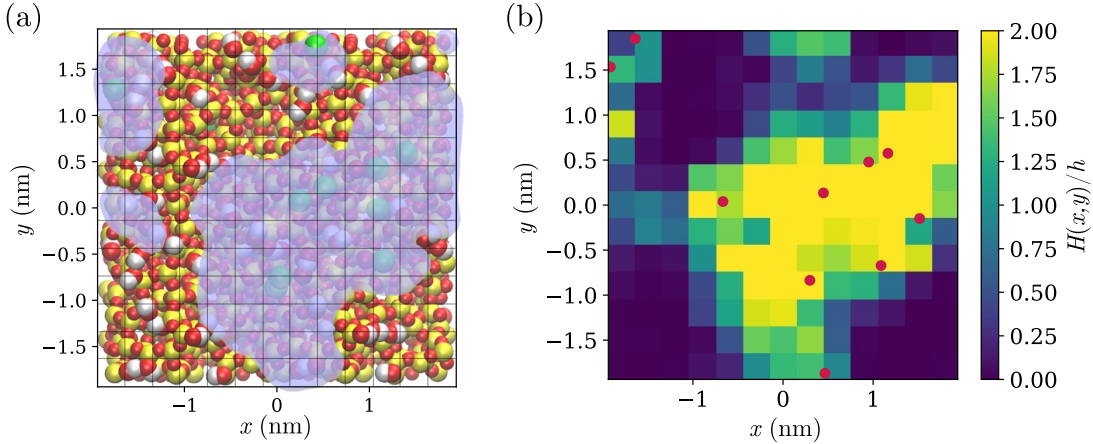

FIG. S14. (a) Simulation snapshot of a water film of thickness $h = 0.27\,{\rm nm}$ ($N_{\rm w} = 162$ and 10 Na$^+$) with a grid of element size $\delta x \simeq \delta y \simeq 0.308\,{\rm nm}$ (green spheres are sodium cations, yellow, red and white spheres are respectively silicon, oxygen and hydrogen atoms of silica substrate, and the transparent blue shape corresponds to water) and (b) the liquid local thickness $H(x, y)$ normalized by $h$ for a similar grid. The red points in (b) correspond to the averaged position of ionized silanol groups. The value within each element is computed every 2000 timesteps and is averaged over time.

Figure S15(a) displays the time-averaged distribution of the number of water molecules within each grid element for four different water film thicknesses. The plot reveals that as the film thickness increases, a larger number of water molecules occupy each grid element, this is a trivial effect. Conversely, in thinner films, there is a higher probability of finding grid elements with no water molecules, indicating the presence of dry spots. Additionally, figure S14(b) illustrates the occupation of water molecules and ionized silanol groups at the same time. It is evident that water molecules preferentially reside in ionized silanol groups. Specifically, for a water film thickness of $h = 0.11$, nm (corresponding to about 65 water molecules), only 15 % of ionized silanol groups SiO$^-$ are found in dry spots. Moreover,

figure S14(b) depicts the positions of the ionized silanol groups (red points), which are predominantly located in wet spots. This observation suggests a preference for water molecules to be present in regions with ionized silanol groups SiO⁻.

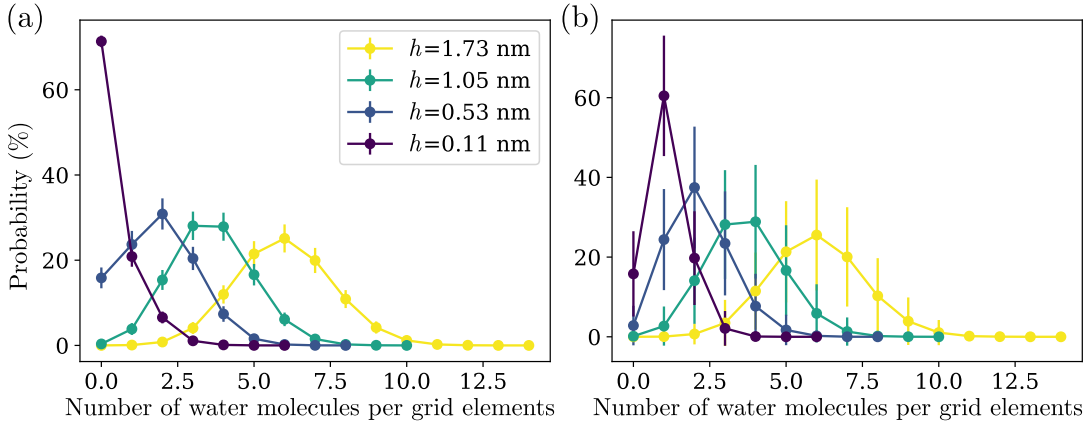

FIG. S15. (a) Time-averaged histogram of the number of water molecules and (b) number of water molecules in the elements $(x_i, y_i)$ containing ionized silanol groups SiO⁻ for 4 different water film thicknesses $h$ and for the system with Na⁺ ions. Results are equivalent for other species.

This water organization appears to be different from what is expected from the theory, and confined experiments in nanochannels [24–26]. Going further, it is possible to quantify the proportion of these dry spots as a function of the water film thickness.

## B. Dry surface ratio

In this section, the dry surface ratio is quantified using the grid and water mass density data presented in the previous section.

Figure S16 illustrates the dry surface ratio as a function of the water film thickness $h$ for the three different cations (filled red circles, light blue triangles and green squares respectively for Na⁺, K⁺ and Li⁺). It is observed that the first dry spot appears at around $h = 0.7\,\text{nm}$ for all considered ions, and the dry spot ratio increases to about $65\,\%$ for the thinnest water film ($h = 0.11\,\text{nm}$). Additionally, the behavior of these measurements is consistent for every cation.

To investigate whether the configuration of the wetting film is influenced by the presence of ionized silanol groups and cations within the film, simulations were conducted without

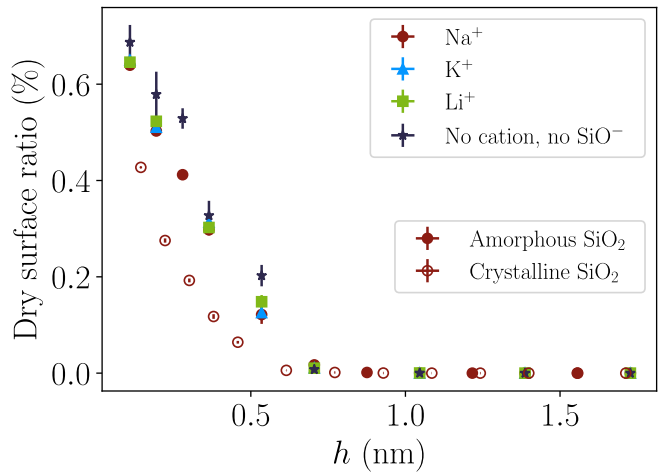

FIG. S16. Dry spot ratio for Na$^+$, K$^+$ and Li$^+$ systems on amorphous SiO$_2$ substrate (filled markers) and Na$^+$ system on crystalline SiO$_2$ (open markers). An equivalent investigation has been carried out for a system composed of water molecules only on the same amorphous SiO$_2$ interface without ionized silanol groups (dark blue star data).

SiO$^-$ and cations on the amorphous silica interface. The results, depicted with dark blue star markers in figure S16, exhibit a behavior similar to the previous measurements. This suggests that the overall structure of the water film is not significantly affected by the presence of ionized silanol groups in this specific system. This latter fact is not completely trivial as it seems that wet spot location do correlate with ionized groups: SiO$^-$ groups thus pin the location of the wetter areas but do not set either their extent nor their limit of appearance. However, it has been previously shown that the wettability of silica does depend on the number of silanol groups present on the surface [14, 27].

Interestingly, the appearance of the first dry spot on the amorphous silica interface occurs at a thickness approximately 2.5 times the size of a water molecule $d_{H_2O}$. As shown in the simulation snapshot in Figure S14(a), the water molecules do not form a uniform, single-layer adsorbed film that covers the entire interface. However, the extent of this behavior depends on the type of interface considered. Figure S16 also compares the amorphous and crystalline silica interfaces for Na$^+$. It can be observed that the amplitude of the dry surface ratio is lower for the flat crystalline interface compared to the rougher amorphous interface. Additionally, the appearance of the first dry spot occurs at a thickness of around 0.6 nm, which is lower than the roughness of the amorphous interface. This suggests that

the roughness of the interface has some influence on the water film structure.

---

soft interface dominated systems: A comparison of phase field and Gibbs dividing surface models.



\end{document}